\documentclass[nofootinbib,aps,prd,groupedaddress,preprintnumbers,%
  showpacs,showkeys,floatfix,amssymb,amsfonts]{revtex4-1}  
\usepackage{hyperref}
\usepackage[usenames]{color}
\usepackage{todonotes}
\usepackage{bbold}
\usepackage{amsmath}
\usepackage{multirow}
\usepackage{graphicx}
\usepackage{xfrac}
\usepackage[utf8]{inputenc}

\usepackage{amsfonts}
\usepackage{amssymb}
\usepackage{extarrows}

\newcommand{\be}{\begin{equation}}
\newcommand{\ee}{\end{equation}}
\newcommand{\bea}{\begin{eqnarray}}
\newcommand{\eea}{\end{eqnarray}}

\newcommand{\meff}{m_{\rm{eff}}}

\newcommand{\avgx}{\langle x \rangle}
\newcommand{\avgxx}{\langle x^2 \rangle}
\newcommand{\Dlr}{\buildrel \leftrightarrow \over D\raise-1pt\hbox{}}

\newcommand{\csw}{c_\mathrm{sw}}

\begin{document}

\title{The Mellin moments $\langle x \rangle$ and $\langle x^2 \rangle$ for the pion and kaon from lattice QCD}

\author{
\vspace*{0.35cm}
  Constantia Alexandrou$^{1,2}$,
  Simone Bacchio$^{1,2}$,
  Ian Clo\"et$^3$,
  Martha Constantinou$^{4}$,\\[1ex]
Kyriakos Hadjiyiannakou$^{1,2}$,
  Giannis Koutsou$^{2}$,
  Colin Lauer$^{3,4}$\\[2ex]
 (ETM Collaboration)
}

\affiliation{
  \vskip 0.25cm
  $^1$Department of Physics, University of Cyprus,  P.O. Box 20537,  1678 Nicosia, Cyprus\\
  \vskip 0.05cm
    $^2$Computation-based Science and Technology Research Center,
  The Cyprus Institute, 20 Kavafi Str., Nicosia 2121, Cyprus \\
  \vskip 0.05cm
  $^3$ Physics Division, Argonne National Laboratory, Argonne, IL\\
  \vskip 0.05cm
  $^4$Department of Physics, Temple University, 1925 N. 12th Street, Philadelphia, PA 19122-1801, USA\\ 
  \vskip 0.05cm
\phantom{-}
\centerline{\today}
 {\includegraphics[scale=0.2]{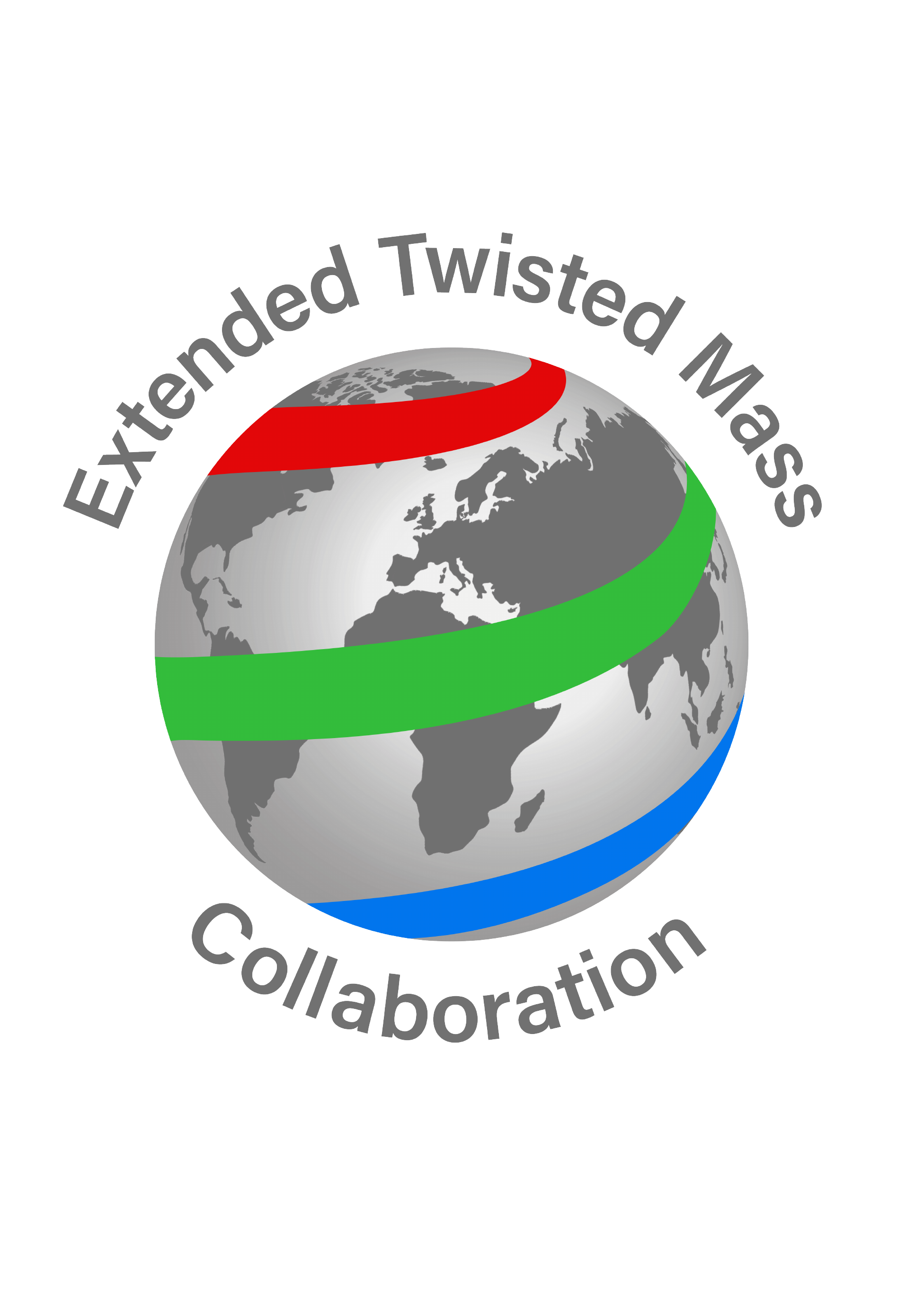}}
  }

\begin{abstract}
We present a calculation of the pion quark momentum fraction, $\langle x \rangle$, and its third Mellin moment $\langle x^2 \rangle$. We also obtain directly, for the first time, $\langle x \rangle$ and$ \langle x^2 \rangle$ for the kaon using local operators. We use an ensemble of two degenerate light, a strange and a charm quark ($N_f=2+1+1$) of maximally twisted mass fermions with clover improvement. The quark masses are chosen so that they reproduce a pion mass of about $260$ MeV, and a kaon mass of 530 MeV. The lattice spacing of the ensemble is 0.093 fm and the lattice has a spatial extent of 3~fm. We analyze several values of the source-sink time separation within the range of $1.12-2.23$~fm to study and eliminate excited-states contributions. The necessary renormalization functions are calculated non-perturbatively in the RI$'$ scheme, and are converted to the $\overline{\rm MS}$ scheme at a scale of 2 GeV. The final values for the momentum fraction are $\langle x \rangle^\pi_{u^+}=0.261(3)_{\rm stat}(6)_{\rm syst}$, $\langle x \rangle^K_{u^+}=0.246(2)_{\rm stat}(2)_{\rm syst}$, and $\langle x \rangle^K_{s^+}=0.317(2)_{\rm stat}(1)_{\rm syst}$. For the third Mellin moments we find $\langle x^2 \rangle^\pi_{u^+}=0.082(21)_{\rm stat}(17)_{\rm syst}$, $\langle x^2 \rangle^K_{u^+}=0.093(5)_{\rm stat}(3)_{\rm syst}$, and $\langle x^2 \rangle^K_{s^+}=0.134(5)_{\rm stat}(2)_{\rm syst}$. The reported systematic uncertainties are due to excited-state contamination. We also give the ratio $\langle x^2 \rangle/\langle x \rangle$ which is an indication of how quickly the PDFs lose support at large $x$.
\end{abstract}

\maketitle

\section{Introduction}
\label{sec:intro}

The pion and kaon provide a good laboratory for studying QCD dynamics at hadronic scales.  Moments of parton distribution functions (PDFs) are important quantities for the study of the internal structure of hadrons. They can directly be computed non-perturbatively within lattice QCD, using local operators, up to $\langle x^3 \rangle$, and yield important insights that complement experimental programs that mostly  measure PDFs for the nucleon.
As the global initiatives to study PDFs in a variety of high-energy processes, such as deep-inelastic lepton scattering (DIS) and Drell-Yan in hadron-hadron collisions, at facilities such as Jefferson Lab, RHIC, Fermilab, and the LHC, intensify, providing theoretical insights on the moments has become very timely. In particular, studying the the pion and kaon moments will provide valuable information for the experimental program of the future Electron-Ion Collider~\cite{NAP25171,Aguilar:2019teb}.

PDFs provide a complementary picture of the structure of hadrons, as compared to electromagnetic form factors. However, unlike form factors PDFs are light-cone dominated quantities, and thus, cannot be calculated directly on a Euclidean lattice. {There exist in the literature alternative approaches to obtain the $x$-dependence of distribution functions, such as the hadronic tensor~\cite{Liu:1993cv}, OPE without an OPE~\cite{Kennedy:1998cu}, auxiliary quark field approaches~\cite{Detmold:2005gg,Braun:2007wv}, quasi-distributions~\cite{Ji:2013dva,Ji:2014gla}, pseudo-distributions~\cite{Radyushkin:2016hsy}, an approach that uses the Compton amplitude~\cite{Chambers:2017dov}, and the current-current correlators method~\cite{Ma:2014jla,Ma:2014jga,Ma:2017pxb}. These methods progress in parallel with direct calculations of moments of PDFs using local operators. For a recent review on these approaches see Refs.~\cite{Cichy:2018mum,Constantinou:2020pek}. Of particular interest are the calculations of the pion and kaon PDFs using some of these methods~\cite{Karpie:2018zaz,Sufian:2019bol,Izubuchi:2019lyk,Joo:2019bzr,Bali:2019dqc,Lin:2020ssv,Gao:2020ito}, some of which have extracted moments of these distributions. Despite the significant progress, dedicated studied are needed to understand the various sources of systematic uncertainties related to these methods. For example, all of these suffer from the ill-defined inverse problem due to the limited number of lattice data entering the reconstruction of the $x$-dependence~\cite{Karpie:2019eiq}. Therefore, having results for the moments of PDFs extracted directly from local operators is imperative and can be used to compare with other studies. The moments are interesting in their own right, as they are extracted from phenomenological analyses of experimental data, enabling direct comparison.}

While the proton has been extensively investigated using lattice QCD, there are only limited studies for pion and kaon structure~\cite{Badier:1983mj,Betev:1985pf,Amendolia:1986wj,Conway:1989fs,Aaron:2010ab}. A few studies exist beyond lattice QCD mainly within models, such as the Dyson-Schwinger equations~\cite{Chang:2013pq,Chang:2013nia,Shi:2018zqd,Bednar:2018mtf} and Nambu--Jona-Lasinio~\cite{Hutauruk:2016sug,Ninomiya:2014kja}. In addition, lattice QCD calculations mostly focus on the pion electromagnetic form factor~\cite{Chambers:2017tuf,Koponen:2017xmw,Alexandrou:2017blh,Wang:2018pii,Owen:2015gva} and the pion average momentum fraction~\cite{Brommel:2006zz,Brommel:2006ww,Bali:2013gya,Abdel-Rehim:2015owa,Alexandrou:2017blh,Oehm:2018jvm}. Given the relatively small amount of experimental data to date it is important to obtain results on the Mellin moments  from \textit{first principle} calculations. 

Pion and kaon structure is relevant to a number of important questions, such as, how hadron masses are generated, the dynamics of chiral symmetry breaking, and the role of pions in nucleon-nucleon interactions. The constrast between the nucleon, and the pion and kaon, is crucial to understand the Standard Model mechanisms that produce hadron masses. For example, in the chiral limit the masses of the pion and kaon vanish, whereas, the nucleon still has a mass on the order of 1\,GeV. As such, the trace anomaly must vanish in the pion/kaon in the chiral limit but is non-vanishing in the nucleon. The pion, the lightest hadronic state in the QCD spectrum, is relevant for chiral symmetry breaking involved in nucleon-nucleon interactions. Pion cloud for instance, can explain why there are more $\overline{d}$ than $\overline{u}$ anti-quarks in the proton sea~\cite{Thomas:2000ny,Chen:2001et,Chen:2001pva,Salamu:2014pka}. Comparisons between pion and kaon structure can reveal interesting aspects of QCD dynamics. For example, a model calculation~\cite{Hutauruk:2016sug} suggests that the strange contribution to the kaon form factors drops faster with increasing momentum transfer, compared to the up quark form factor, which has been interpreted as a consequence of confinement. 

The rest of the paper is organized as follows: In Sec.~\ref{sec2} we present the theoretical setup of the calculation and  the appropriate decomposition to obtain $\avgx$ and $\avgxx$ for mesons. In Sec.~\ref{sec3a} we provide the details of the lattice formulation, the parameters of the ensemble employed, as well as the calculation of the correlation functions needed in this work. The analysis for the extraction of reliable estimates for the non-perturbative renormalization functions is  described in Sec.~\ref{sec:renorm}, as well as the details of the non-perturbative prescription. The various analyses on the two-point correlation functions for extracting the pion and kaon masses corresponding to the ensemble under study are presented in Sec.~\ref{sec4}. In the same section we also include a thorough investigation of excited states for both $\avgx$ and $\avgxx$, as well as an alternative kinematic setup for extracting $\avgx$. The final results along with comparison with other studies and phenomenology are presented in Sec.~\ref{sec5}. In Sec.~\ref{sec6} we summarize our work and conclude.

\section{Theoretical setup}
\label{sec2}

To calculate $\langle x \rangle$, we compute the  meson matrix elements $\langle M(p^\prime)|{\cal O}^{\{\mu\nu\}}|M(p)\rangle$ for $p^\prime=p$, where ${\cal O}^{\{\mu\nu\}}$ is the one-derivative vector operator defined as
\begin{equation}
{\cal O}^{\{\mu\nu\}}\equiv \overline{\psi} \left[ \frac{1}{2}\left( \gamma^\mu \Dlr^\nu +\gamma^\nu \Dlr^\mu \right)  - \frac{1}{4}\sum_{\rho=1}^{4}\delta_{\mu\nu}\gamma^\rho\Dlr^\rho\right]\psi\,.
\label{eq:Op_x}
\end{equation}
where the notation $\{ \cdots\}$ means symmetrization and traceless,
with
$\Dlr=\frac{1}{2}\left( \stackrel{\rightarrow}{D}_\mu-\stackrel{\leftarrow}{D}_\mu\right)$, $\stackrel{\rightarrow}{D}_\mu=\frac{1}{2}\left(\nabla_\mu +\nabla^*_\mu \right)$ \cite{PhysRevD.93.039904}, and $\psi$ corresponds to the up or down quark for the pion, and the up or strange quark for the kaon. $\{...\}$ indicates symmetrization over the indices, in this case $\mu$ and $\nu$, as well as subtraction of the trace, to avoid mixing with other operators~\cite{Capitani:1994qn,Beccarini:1995iv,Gockeler:1996mu}. 

The higher moment $\langle x^2 \rangle$ is accessed using a fermion operator with two covariant derivatives, ${\cal O}^{\mu\nu\rho} = \overline{\psi} \gamma^{\mu} D^{\nu} D^{\rho}\, {\psi}$. To avoid any mixing, we choose the indices $\mu,\,\nu,\,\rho$ to be different~\cite{Beccarini:1995iv,Capitani:2000wi,Capitani:2000aq}. Therefore, only a symmetrization over these indices is needed, that is,
\begin{equation}
{\cal O}^{\{\mu\nu\rho\}} \equiv \frac{1}{6} \Bigl(
{\cal O}^{\mu\nu\rho} + 
{\cal O}^{\mu\rho\nu} + 
{\cal O}^{\nu\mu\rho} + 
{\cal O}^{\nu\rho\mu} + 
{\cal O}^{\rho\mu\nu} +  
{\cal O}^{\rho\nu\mu} \Bigr) \,.
\label{eq:Op_x2}
\end{equation}

The meson matrix elements decompose into two generalized form factors, $A_{20}$ and $B_{20}$ for the one-derivative vector operator, and $A_{30}$ and $B_{30}$ for the two-derivative operator. The kinematic coefficients in Euclidean space are given by:
\begin{eqnarray}
\label{decomp_x}
\langle M(p') |{\cal O}^{\{\mu\nu\}} | M(p) \rangle &=& C\,\left[ 2 P^{\{\mu} P^{\nu\}} \,A_{20}(Q^2) + 2 \Delta^{\{\mu} \Delta^{\nu\}} \,B_{20}(Q^2) \right]\,,\\[1ex]
\langle M(p') | {\cal O}^{\{\mu\nu\rho\}}  | M(p) \rangle &=& C\,\left[ 2 i\,P^{\{\mu} P^{\nu}  P^{\rho\}}\,A_{30}(Q^2) + 2 i\, \Delta^{\{\mu} \Delta^{\nu} P^{\rho\}}\,B_{30}(Q^2)\right]\,.
\label{decomp_x2}
\end{eqnarray}
In the above decompositions, $P$ is the average of the initial and final momenta of the meson, $P=(p+p')/2$, and $\Delta$  their difference,  $\Delta=p'-p$. $Q^2$ is the momentum transferred squared. $C$ is a kinematic factor related to the normalization of the two-point functions. Therefore, $C$ depends on the frame employed and the momentum transferred. Based on our conventions, we obtain $C=\frac{1}{\sqrt{4 E(p) E(p')}}$ for a general frame. $m_M$ is the mass of  meson $M$ and $E(p){=}\sqrt{m_M^2 + \vec{p}\,^2}$ is the energy at momentum $\vec{p}$. Therefore, $C$ does not depend on the spatial directions of the momentum, only on $\vec{p}\,^2$. Note also, that the generalized form factors $A_{i0}$ and $B_{i0}$ are functions of the momentum transferred squared, and are independent of the kinematic setup.

The quantities of interest are obtained from the forward limit of the matrix elements, leading to $\avgx\equiv A_{20}(0)$, $\avgxx\equiv A_{30}(0)$. The decomposition of Eq.~(\ref{decomp_x}) takes a simple form for mesons at rest, and in fact, at $Q^2=0$ there is only one matrix element contributing, which has $\mu=\nu=4$ where we denote by index 4 the temporal direction. In such a simplified case, Eq.~(\ref{decomp_x}) becomes
\begin{eqnarray}
    \langle M(0)|{\cal O}^{44}|M(0)\rangle =  -\frac{3 m_M}{4}  \avgx^M\,.
    \label{decomp_x_simple}
\end{eqnarray}
The index ``$M$'' indicates the meson of interest. The kinematic coefficient of $A_{30}$ becomes zero in the rest frame ($\vec{p}\,'=\vec{p}=0$) and in the forward limit, unless all the indices of the operator are temporal. This is is not an optimal option, as ${\cal O}^{444}$ suffers from mixing with lower-dimension operators~\cite{Capitani:1994qn,Beccarini:1995iv,Gockeler:1996mu}, making the extraction of $\avgxx$ unreliable. In fact, to eliminate mixing even with operators of equal dimension, all indices must be different from each other, which is the choice we employ in this work. Given these constraints, the only way to obtain $\avgxx$ is to work in a frame in which the meson is moving with some momentum $\vec{p}\,'=\vec{p}\ne0$ (boosted frame). In the forward limit, the momentum is the same at the source and the sink, $p'=p=(i E,p_x,p_y,p_z)$. For $\mu\ne\nu\ne\rho\ne\mu$, all spatial components of the momentum must be non-zero to extract $\avgxx$ directly from lattice data, without the need of applying fits on matrix elements with finite momentum transfer. In the boosted frame, the matrix elements are related to their corresponding Mellin moments via: 
\begin{eqnarray}
\label{decomp_x_2}
    \langle M(p)|{\cal O}^{44}|M(p)\rangle &=& \frac{1}{2 E_M(p)} \left(\frac{m_M^2}{2}-2 (E_M(p))^2\right) \avgx^M\,,\\[1ex]
        \langle M(p)|{\cal O}^{\mu\nu4}|M(p)\rangle &=&  - p_\mu\,p_\nu   \avgxx^M\,,
    \label{decomp_x2_2}
\end{eqnarray}
which include all kinematic factors and normalizations. In Eq.~(\ref{decomp_x2_2}) we take one of the indices to be temporal, which simplifies the kinematic factor of $\langle x^2 \rangle$. {The other two indices are spatial and different from each other. Therefore, there are three different combinations of operators: $O^{\{xy4\}}$, $O^{\{xz4\}}$, and $O^{\{yz4\}}$, each symmetrized over its indices. In this calculation, we extract all these combinations and we average using their corresponding kinematic factor shown in Eq.~(\ref{decomp_x2_2}).}

\section{Lattice Setup}
\label{sec3}

\subsection{Lattice Action}
\label{sec3a}

In this work we employ one ensemble~\cite{Alexandrou:2018egz} of $N_f=2+1+1$ twisted mass fermions with a clover term and the Iwasaki improved gluon action. The ensemble is generated by the Extended Twisted Mass Collaboration (ETMC). The fermionic part of the action is written in the physical basis as
\begin{equation}
    \label{eq:Dtm}
    S[\psi,\bar{\psi},U] = a^4 \sum_x \overline{\psi}(x)\left( D[U] + \mu_q - i\gamma_5\tau_3 \left[W_\mathrm{cr} +
    \frac{i}{4}c_\mathrm{sw}\sigma^{\mu\nu}\mathcal{F}^{\mu\nu}[U]\right]\right)\psi(x)\,.
\end{equation}
$D = \gamma_\mu(\nabla^\ast_\mu+\nabla_\mu)/2$, $\nabla_\mu$ and $\nabla_\mu^\ast$ are the forward and backward lattice covariant derivatives, and $\mu_q$ is the twisted quark mass~\cite{Oehm:2018jvm}. $W_\mathrm{cr} = -(a/2)\nabla_\mu^\ast\nabla_\mu + m_\mathrm{cr}$ and $m_\mathrm{cr}$ is the bare untwisted mass tuned to its critical value, which gives automatic $\mathcal{O}(a)$ improvement~\cite{Frezzotti:2000nk}, requiring no further improvements on the operator level. The last term is the clover term multiplied by the Sheikoleslami-Wohlert improvement coefficient $\csw$. Since we achieve ${\cal O}(a)$ improvement from the critical mass, $\csw$ is used to reduce isospin symmetry breaking effects~\cite{PhysRevD.74.034501}. Other parameters for this ensemble are $\kappa=0.1400645$ $c_\mathrm{sw}=1.74$, $\mu_{l}=0.003$, $\mu_\Sigma=0.1408$, $\mu_\Delta=0.1521$. The remaining parameters of the simulation are given in Table \ref{tab:params}. 

 \begin{table}[h!]
\centering
\renewcommand{\arraystretch}{1.2}
\renewcommand{\tabcolsep}{6pt}
  \begin{tabular}{| l| c | c | c | c | c  | c | c |}
  \hline
    \multicolumn{8}{|c|}{Parameters} \\
    \hline
 Ensemble   & $\beta$ & $a$ [fm] & volume $L^3\times T$ & $N_f$ & $m_\pi$ [MeV] &
$L m_\pi$ & $L$ [fm]\\
    \hline
cA211.32 & 1.726 & 0.093  & $32^3\times 64$  & $u, d, s, c$ & 260
& 4 & 3.0 \\
    \hline
    \end{tabular}
  \caption{Parameters of the ensemble used in this work.}
  \label{tab:params}
  \vspace*{0.2cm}
  \end{table}
For the interpolating fields, $J_M$, of  the mesons under study we take
\bea
J_{\pi^+} = \overline{d}\gamma_5 u\,, \\
J_{K^+} = \overline{s}\gamma_5 u\,.
\eea
A useful consequence of the pseudoscalar structure of the pion, as well as the $\gamma_5$-hermiticity relation of the twisted mass quark propagators
\begin{equation}
    G_{u}(x,x')=\gamma_5 G_{d}^\dagger(x',x)\gamma_5\,,
\end{equation}
is that we only need to calculate the up quark contribution to the pion three-point functions. 
 The pion and kaon interpolating fields are smeared using Gaussian smearing at both  source and sink. The smearing parameters are tuned separately for the pion and kaon. We  use the same value of $\alpha_G$ but varying the number of smear iterations $N_G$ for the light and strange quarks. An optimal choice for $N_G$ is based on the criterion that the root mean squared radius of the smeared source reproduces the experimental radius of the pion~\cite{AMENDOLIA1984116} for the light quarks and the experimental radius of the kaon~\cite{AMENDOLIA1986435} for the strange quarks. In this work we obtain $(\alpha_{G},N_{G})=(0.2,50)$ for the light quarks and $(\alpha_{G},N_{G})=(0.2,40)$ for the strange quark.  APE smearing is applied on the gauge links that enter the gaussian smearing with parameters ($\alpha_{APE},N_{APE})=(0.5,50)$.

\begin{figure}[h]
    \centering
    \includegraphics[scale=0.25]{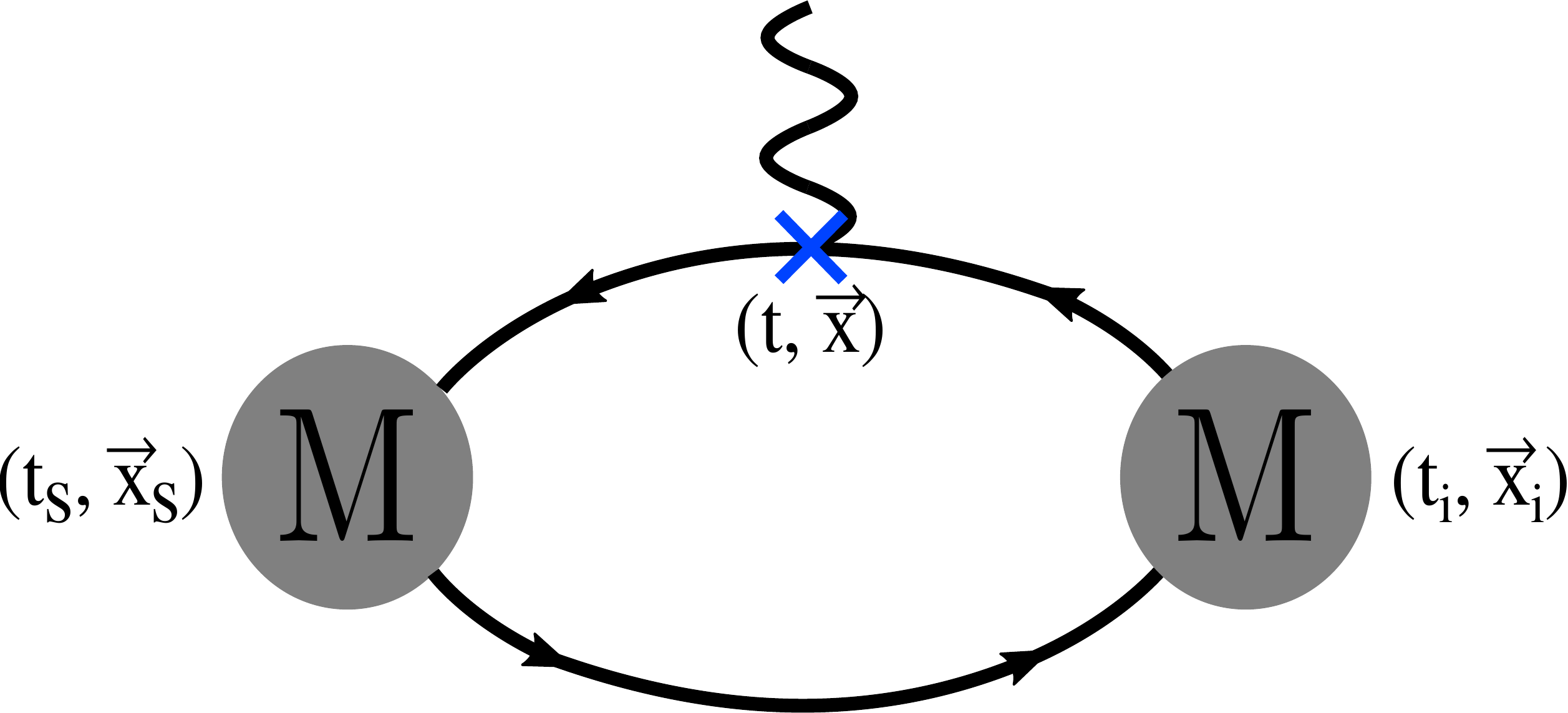}
\caption{Connected diagram for the three-point function entering the calculation of $\langle x \rangle$ and $\avgxx$. The wavy line corresponds to the operator insertion.}
\label{fig:3ptDiagram}
\end{figure}
We study the connected contribution to the matrix elements of ${\cal O}^{44}$ and ${\cal O}^{\mu\nu4}$, which is shown in Fig.~\ref{fig:3ptDiagram}. For the calculation of the three-point functions we use the fixed sink sequential inversion approach. The three-point correlation functions are calculated at zero momentum transfer,
\begin{equation}
C_M^\Gamma(t,t_s,\vec{p}) = \sum_{\vec{x}_s,\vec{x}} J_M(t_s,\vec{x}_s){\cal O}^{\mu\nu}(t,\vec{x})J^\dag_M(t_i,\vec{x}_i) e^{-i\vec{p}\cdot(\vec{x}_s-\vec{x}_i)}\,,
\end{equation} 
where $t_i$, $t$, $t_s$ are the source, insertion and sink Euclidean times, respectively. The corresponding spatial coordinates of the source, current insertion and sink are $\vec{x}_i$, $\vec{x}$, $\vec{x}_s$. Without loss of generality we will take the source to be at $t_i=0$, so that the source-sink separation $t_s -t_i = t_s$. 
For a general insertion current ${\cal O}_\Gamma=\bar{u}\Gamma u \pm \bar{d}\Gamma d$, the three-point functions can be written in terms of the up and down parts
\be
C_M^{\mu\nu}(t,t_s) = \sum_{\vec{x}_s,\vec{x}}  J_M(t_s,\vec{x}_s)[\bar{u}\Gamma(t,\vec{x}) u \pm \bar{d}\Gamma(t,\vec{x}) d]J^\dag_M(0,\vec{0}) =C^{\Gamma}_{M,u}(t,t_s)\pm C^{\Gamma}_{M,d}(t,t_s)\,.
\ee
Performing the Wick contractions for the pion, and applying the $\gamma_5$-hermiticity, it can be shown that, for $\pi^+$, the up and down parts are related by
\begin{equation}
    C^{\Gamma}_{\pi,d}=\pm \left (C^{\Gamma}_{\pi,u}\right)^*\,.
    \label{eq:pion_conj}
\end{equation}
The plus / minus sign comes from the fact that a general $\gamma$-structure is either $\gamma_5$-hermitian or anti-$\gamma_5$-hermitian, i.e. $\gamma_5 \Gamma^\dagger \gamma_5 = \pm \Gamma$. Both the one-derivative vector and two-derivative vector operators are $\gamma_5$-hermitian, and therefore,
\begin{gather}
\label{eq:pion_u_d_x}
C^{44}_{\pi,u+d}=2C^{44}_{\pi,u}\,, \\[1ex]
C^{\mu\nu4}_{\pi,u+d}=2C^{\mu\nu4}_{\pi,u}\,.
\label{eq:pion_u_d_x2}
\end{gather}
In the results presented here, we focus on the $u^+$ contribution to the pion, where the $q^+\equiv q+\bar{q}$ notation has been adopted. Note that, based on Eqs.~(\ref{eq:pion_u_d_x}) and (\ref{eq:pion_u_d_x2}), $\avgx^\pi_{u^+}=\avgx^\pi_{{d}^+}$ and $\avgxx^\pi_{u^+}=\avgxx^\pi_{{d}^+}$, that is,
$\avgx^\pi_{u^+ + {d}^+}=2 \avgx^\pi_{u^+}$ and $\avgxx^\pi_{u^+ + {d}^+}=2 \avgxx^\pi_{u^+}$. This discussion is relevant to the comparison with phenomenological results presented in Sec.~\ref{sec6}. We note that Eqs.~\ref{eq:pion_conj}, \ref{eq:pion_u_d_x}, \ref{eq:pion_u_d_x2} are only applicable for the pion case, whereas for the kaon due to different mass of the light and strange quarks such relations do not hold.

We analyze 122 configurations, separated by 20 trajectories to reduce auto-correlation effects. In the rest frame, we use 16 randomly chosen source positions on each configuration, giving a total statistics of 1952. In the boosted frame, we use 32 source position for a total statistics of 3904. For the calculation in the rest frame, we use six source-sink time separations, namely $t_s/a=12,\,14,\,16,\,18,\,20,\,24$, corresponding to $t_s=1.12-2.23$~fm. This allows for a thorough investigation and elimination of possible contributions from excited states. Based on the analysis of the results in the rest frame we concluded that a subset of $t_s/a=14,\,16,\,18$ is sufficient for extracting the ground state matrix elements. Thus, we only use these three time separations for the  computation of $\avgx$ and $\avgxx$, in the boosted frame. 

According to the decomposition of Eq.~(\ref{decomp_x_2}), $\avgxx$ can be obtained using momentum boost with at least two non-zero spatial components, with the lowest momentum being $\vec{p}_i=\frac{2\pi}{L}(\pm1,\pm1,0)$ (12 combinations). 
In this work, we employ, for $\avgxx$, momenta of the class $\vec{p}_i\,^2=\frac{12\pi^2}{L^2}$, which corresponds to 8 combinations for the spatial components, $\vec{p}_i=\frac{2\pi}{L} (\pm1,\pm1,\pm1)$. With the same setup we also obtain $\avgx$, for a qualitative comparison with the rest frame, and the scaling of the statistical uncertainties. The choice $\vec{p}_i\,^2=\frac{12\pi^2}{L^2}$ is optimal for two reasons: While it increases the statistical uncertainties as compared to momenta $\vec{p}_i\,^2=\frac{8\pi^2}{L^2}$, the computational cost for the same number of configurations is reduced by $33\%$ due to the smaller number of permutations. Also, the class  $\vec{p}_i\,^2=\frac{12\pi^2}{L^2}$ allows one to access, with the same setup, other quantities, such as $\langle x^3 \rangle$, as well as form factors and generalized form factors. These quantities will be presented in a follow-up work.

\subsection{Renormalization}
\label{sec:renorm}

The renormalization of the bare matrix elements is multiplicative, and the renormalization functions are calculated non-perturbatively using the Rome-Southampton method (RI$'$ scheme)~\cite{Martinelli:1994ty}. 
The estimates are converted to the $\overline{\rm MS}$-scheme and evolved at a renormalization scale of $\overline{\mu}=$2 GeV. We refer to the renormalization function of ${\cal O}^{\mu\mu}$ and ${\cal O}^{\mu\nu\rho}$ ($\mu\ne\nu\ne\rho\ne\mu$) as $Z_{\rm vD}$ and $Z_{\rm vDD}$, respectively.
The renormalization function in the RI$'$ scheme are determined by the conditions
\be
   Z_q = \frac{1}{12} {\rm Tr} \left[(S^L(p))^{-1}\, S^{{\rm Born}}(p)\right] \Bigr|_{p^2=\mu_0^2}\,,\qquad
   Z_q^{-1}\,Z_{\cal O}\,\frac{1}{12} {\rm Tr} \left[\Gamma_{\cal O}^L(p)
     \,\left(\Gamma_{\cal O}^{{\rm Born}}(p)\right)^{-1}\right] \Bigr|_{p^2=\mu_0^2} = 1\, ,
\label{renormalization cond}
\ee
where $p$ is the momentum of the vertex function, set to the RI$'$ renormalization scale, $\mu_0$. $S^{{\rm Born}}$ ($\Gamma_{\cal O}^{{\rm Born}}$) is the tree-level value of the fermion propagator (operator), and the trace is taken over spin and color. We use the momentum source method~\cite{Gockeler:1998ye}, which is successfully employed for twisted mass fermions~\cite{Alexandrou:2010me,Alexandrou:2012mt,Alexandrou:2015sea}. This method achieves per mil accuracy even with a small number of configurations. In the results presented here we use 10 configurations. In order to reduce discretization effects we use momenta that have the same spatial components, that is:
\begin{equation}
(a p) \equiv 2\pi \left(\frac{n_t}{L_t}+\frac{1}{2\,L_t},
\frac{n_x}{L_s},\frac{n_x}{L_s},\frac{n_x}{L_s}\right)\,,  \qquad\quad n_t \,\epsilon\, [2, 9]\,,\quad n_x\,\epsilon\, [2, 5]\,,\quad (a p)^2 \in [0.9 - 6.7]\,,
\end{equation} 
where $L_t$ ($L_s$) is the temporal (spatial) extent of the lattice. These momenta are chosen to have suppressed non-Lorentz invariant contributions ($ {\sum_i p_i^4}/{(\sum_i p_i^2 )^2}{<}0.3$), which is based on empirical arguments~\cite{Constantinou:2010gr}. We improve the non-perturbative estimates for $Z_{\rm vD}$ by subtracting finite lattice effects using the procedure outlined in Refs.~\cite{Constantinou:2014fka,Alexandrou:2015sea}. The latter are computed to one loop in perturbation theory and to all orders in the lattice spacing, ${\cal O}(g^2\,a^\infty)$. Such a procedure is not yet available for two-derivative operators. However, we partly improve $Z_{\rm vDD}$, by subtracting the ${\cal O}(g^2\,a^\infty)$ artifacts from $Z_q$ which enters the renormalization condition for $Z_{\rm vDD}$ in Eq.~(\ref{renormalization cond}).

\begin{table}[!h]
\begin{center}
\begin{tabular}{ccc}
\hline
\hline
$\,\,\,$        $\,\,\,$              & $\beta=1.726$, $\,\,\,a=0.093$ fm  &          \\ \hline
\hline
$a \mu$   & $a m_{PS}$ & lattice size\\
\hline
$\,\,\,$  0.0060$\,\,\,$         & 0.1680   & $24^3 \times 48$ \\  
$\,\,\,$  0.0080$\,\,\,$         & 0.1916   & $24^3 \times 48$ \\          
$\,\,\,$  0.0100$\,\,\,$         & 0.2129   & $24^3 \times 48$ \\  
$\,\,\,$  0.0115 $\,\,\,$        & 0.2293   & $24^3 \times 48$ \\ 
$\,\,\,$  0.0130 $\,\,\,$        & 0.2432   & $24^3 \times 48$  \\ 
\hline
\hline
\end{tabular}
\caption{Parameters for the $N_f=4$ ensembles used for the renormalization functions.} 
\label{tab:Z_ensembles}
\end{center}
\end{table}

For a proper chiral extrapolation, we calculate the renormalization functions on several ensembles with all masses of quark flavors degenerate ($N_f=4$). We use five ensembles at different values for the pion mass, which are produced with the same $\beta$ value as the one of the cA211.32 ensemble analysed for the matrix elements. The parameters of the $N_f=4$ ensembles are given in Table~\ref{tab:Z_ensembles}. The chiral limit is taken using a quadratic fit with respect to the pion mass. For both $Z_{\rm vD}$ and $Z_{\rm vDD}$, we find a negligible dependence on the pion mass, as can be seen in Table~\ref{table:Zmpi} for two representative renormalization scales ($(a \mu_0)^2=2,\,4$). On each $N_f=4$ ensemble we use 23 values of the initial RI$'$ scale $\mu_0$ ranging from ($(a\mu)^2 \in [0.9 - 6.7]$). Each value is converted and evolved to $\overline{\rm MS} (2~{\rm GeV}$) using an intermediate Renormalization Group Invariant scheme defined in continuum perturbation theory. A linear fit with respect to $(a \mu_0)^2$ is applied on the $\overline{\rm MS}$ values to eliminate residual dependence on the initial scale $\mu_0$. Such a dependence may be present due to finite-$a$ effects and/or truncation of the conversion factor (to three loops in perturbation theory).

  \begin{table}[h]
  \begin{minipage}{8cm}
  \hspace*{4cm}
    \begin{tabular}{ccc}
      \hline\hline\\[-2.5ex]
   \, &    $\,\,Z_{\rm vD}^{{\rm RI}'}$ & \, \\
            \hline\\[-2.5ex]
 $a m_{PS}$ & $\,\,(a \mu_0)^2{=}2$ & $\,\,(a \mu_0)^2{=}4$ \\[0.5ex]
\hline\\[-2.5ex]
$\,\,\,$0.1680 &$\,\,\,$ 1.1762(2)     &$\,\,\,$ 1.1043(1)  \\
$\,\,\,$0.1916 &$\,\,\,$ 1.1770(3)     &$\,\,\,$ 1.1045(2)    \\
$\,\,\,$0.2129 &$\,\,\,$ 1.1773(2)     &$\,\,\,$ 1.1046(1)  \\
$\,\,\,$0.2293 &$\,\,\,$ 1.1782(2)     &$\,\,\,$ 1.1048(1)  \\
$\,\,\,$0.2432 &$\,\,\,$ 1.1779(2)     &$\,\,\,$ 1.1047(1)  \\ [0.5ex]
      \hline\hline
    \end{tabular}
\end{minipage}
\hfill
  \begin{minipage}{8cm}
    \hspace*{-4cm}
    \begin{tabular}{ccc}
      \hline\hline\\[-2.5ex]
   \, &    $\,\,Z_{\rm vDD}^{{\rm RI}'}$ & \, \\
            \hline\\[-2.5ex]
 $a m_{PS}$ & $\,\,(a \mu_0)^2{=}2$ & $\,\,(a \mu_0)^2{=}4$ \\[0.5ex]
\hline\\[-2.5ex]
$\,\,\,$0.1680 &$\,\,\,$ 1.4870(4)     &$\,\,\,$ 1.3722(2) \\
$\,\,\,$0.1916 &$\,\,\,$ 1.4890(5)     &$\,\,\,$ 1.3733(2)  \\
$\,\,\,$0.2129 &$\,\,\,$ 1.4888(5)     &$\,\,\,$ 1.3732(3) \\
$\,\,\,$0.2293 &$\,\,\,$ 1.4922(4)     &$\,\,\,$ 1.3751(2) \\
$\,\,\,$0.2432 &$\,\,\,$ 1.4914(6)     &$\,\,\,$ 1.3748(3) \\ [0.5ex]
      \hline\hline
    \end{tabular}
\end{minipage}
        \caption{Pion mass dependence of the renormalization function $Z_{\rm vD}$ (left panel) and $Z_{\rm vDD}$ (right panel) in the RI$'$ scheme. The first column is the pion mass (in lattice units) for the ensemble, the second (third) is the renormalization function at scale $(a\mu_0)^2{=}2$ ($(a\mu_0)^2{=}4$). The number in the parenthesis is the statistical error.}\label{table:Zmpi}
  \end{table}

\begin{figure}[!h]
\centerline{\includegraphics[scale=0.43]{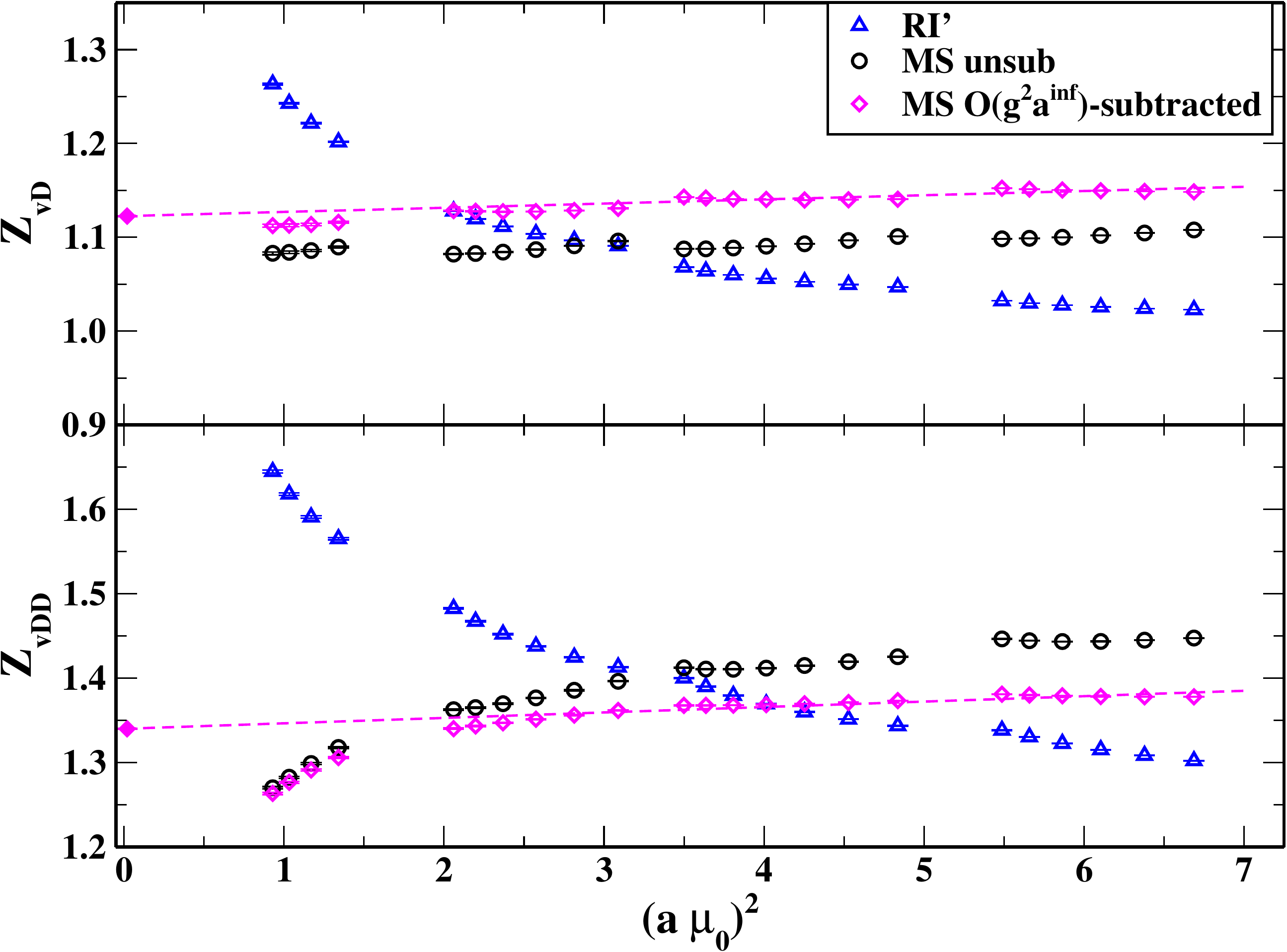}}
\caption{Chirally extrapolated results for $Z_{\rm vD}$ (top) and $Z_{\rm vDD}$ (bottom). Blue triangles correspond to RI$'$ scheme, black circles to ${\overline{\rm MS}}$ scheme and magenta diamonds to the subtracted results in the  ${\overline{\rm MS}}$ scheme. The data are plotted as a function of the initial RI$'$ scale $(a\,\mu_0)^2$. The dashed line corresponds to the fit of Eq.~(\ref{Zfinal}), and the filled magenta diamonds represent our final values for ${\cal Z}_{\rm vD}$ and  ${\cal Z}_{\rm vDD}$.}
\label{ZRIMS}
\end{figure}

\vspace*{0.25cm}
In Fig.~\ref{ZRIMS} we show $Z_{\rm vD}$ and  $Z_{\rm vDD}$ in the RI$'$  and ${\overline{\rm MS}}$ schemes as a function of the initial RI$'$ renormalization scale, $\mu_0$. 
$Z_{\cal O}^{\overline{\rm MS}}$ are given at $\mu=2$ GeV, and the purely non-perturbative data exhibit a small residual dependence on the initial scale $\mu_0$ they were evolved from. A procedure of subtracting the finite-$a$ effects to ${\cal O}(g^2 a^\infty)$ is also applied on $Z_{\rm vD}$. For $Z_{\rm vDD}$ the improvement is only applied to  $Z_q$. We find that for both cases, subtracted results have a much smaller slope than the non-substracted ones, demonstrating the effectiveness of the artifact-subtraction procedure.

We eliminate any residual $(a\mu_0)^2$ dependence in each renormalization function by using the Ansatz
\begin{equation}
Z_{\cal O}(a\,p) = {\cal Z}_{\cal O} + Z_{\cal O}^{(1)}\cdot(a\,\mu_0)^2\,.
\label{Zfinal}
\end{equation}
${\cal Z}_{\cal O}$ corresponds to the final value of the renormalization function for operator ${\cal O}$. We obtain ${\cal Z}_{\rm vD} = 1.123(1)(5)$ and ${\cal Z}_{\rm vDD} = 1.340(1)(15)$, where the numbers in the first and second parentheses are the statistical and systematic errors, respectively. The source of systematics is related to the $(a\,\mu_0)^2\to 0$ extrapolation. The final value uses the fit interval $(a\,\mu_0)^2\, \epsilon\, [2-7]$ and the systematic is estimated by varying the lower range of the fit range {between $(a\,\mu_0)^2){\rm low}\, \epsilon\, [2-4]$.} The reported uncertainty is the difference with the value obtained from $(a\,\mu_0)^2\, \epsilon\, [4-7]$.

\vspace*{0.25cm}

\section{Analysis Methods}
\label{sec4}
\subsection{Effective Mass}

One of the important ingredients in the determination of the Mellin moments is the mass (energy) of the meson in the rest (boosted) frame. This is needed because the ground-state energy enters in the decomposition of Eqs.~(\ref{decomp_x}) - (\ref{decomp_x_2}). We implement two fits for extracting the ground state energy from the two-point correlation functions 
\begin{equation}
    C^{\rm 2pt}_M(t,\vec{p})=\sum_{\vec{x}} J_M(t,\vec{x})J_M^\dag (0,\vec{0}) e^{i\vec{p}\cdot\vec{x}}\,,
\end{equation}
 as described below. We exploit the symmetry properties and we symmetrize the correlator corresponding to $t$ and $T-t$, for  $ t \in [0,T/2]$, i.e., the value at $t$ have been averaged with their corresponding value at $T-t$.

\medskip
\textit{Plateau method:} The first method relies on a single-state fit where the effective mass (energy) is fitted to a  constant with respect to $t$. The fit is taken over a range of $t$ where the effective mass (energy) becomes time independent (plateau region). We calculate the effective mass from the symmetrized two-point function according to
\begin{equation}
    \meff^M(t) = \frac{1}{2}\ln{\left[\frac{C^{\rm 2pt}_M(t-1)+\sqrt{(C^{\rm 2pt}_M(t-1))^2-(C^{\rm 2pt}_M(\frac{T}{2}))^2}}
    {C^{\rm 2pt}_M(t+1)+\sqrt{(C^{\rm 2pt}_M(t+1))^2-(C^{\rm 2pt}_M(\frac{T}{2}))^2}}\,\right]}.
    \label{eq:meff}
\end{equation}
We test several values for the lower value of $t$ entering the plateau fit, $t_{\rm low}/a \in [11-19]$ for the rest frame and $t_{\rm low}/a \in [8-14]$ for the boosted frame, while the maximum value is fixed to $t/a=31$. 

\medskip
\textit{Two-state fit:} The second method is a two-state fit in which we include the first excited state in the fit Ansatz given by
\begin{equation}
    C^{\rm 2pt}_M(t) = c_0 \left(e^{-E_0 t}+e^{-E_0 (T-t)}\right) + c_1 \left(e^{-E_1 t}+e^{-E_1 (T-t)}\right)\,.
    \label{eq:tsf_twop}
\end{equation}
The amplitudes $c_0$ and $c_1$, as well as the ground and first excited state energies $E_0$, $E_1$ are fit parameters.
An alternative procedure is to apply the two-state fit on $\meff$ directly, by substituting Eq.~(\ref{eq:tsf_twop}) into Eq.~(\ref{eq:meff}). This way, one of the amplitudes cancels out and the fit consists of three free parameters. 
Here we employ both procedures to cross-check the consistency of the results.  We note that extracting the amplitude $c_0$ is needed in order to calculate the ratios between the three-point and two-point functions as described in the next section (see, e.g., Eq.~(\ref{eq:ratio})). 
The two-state fit is taken over the range $t \in [t_{\rm low}-31a]$, with $t_{\rm low}/a \in [1-4]$.

\begin{figure}[h!]
    \centering
    \includegraphics[scale=0.45]{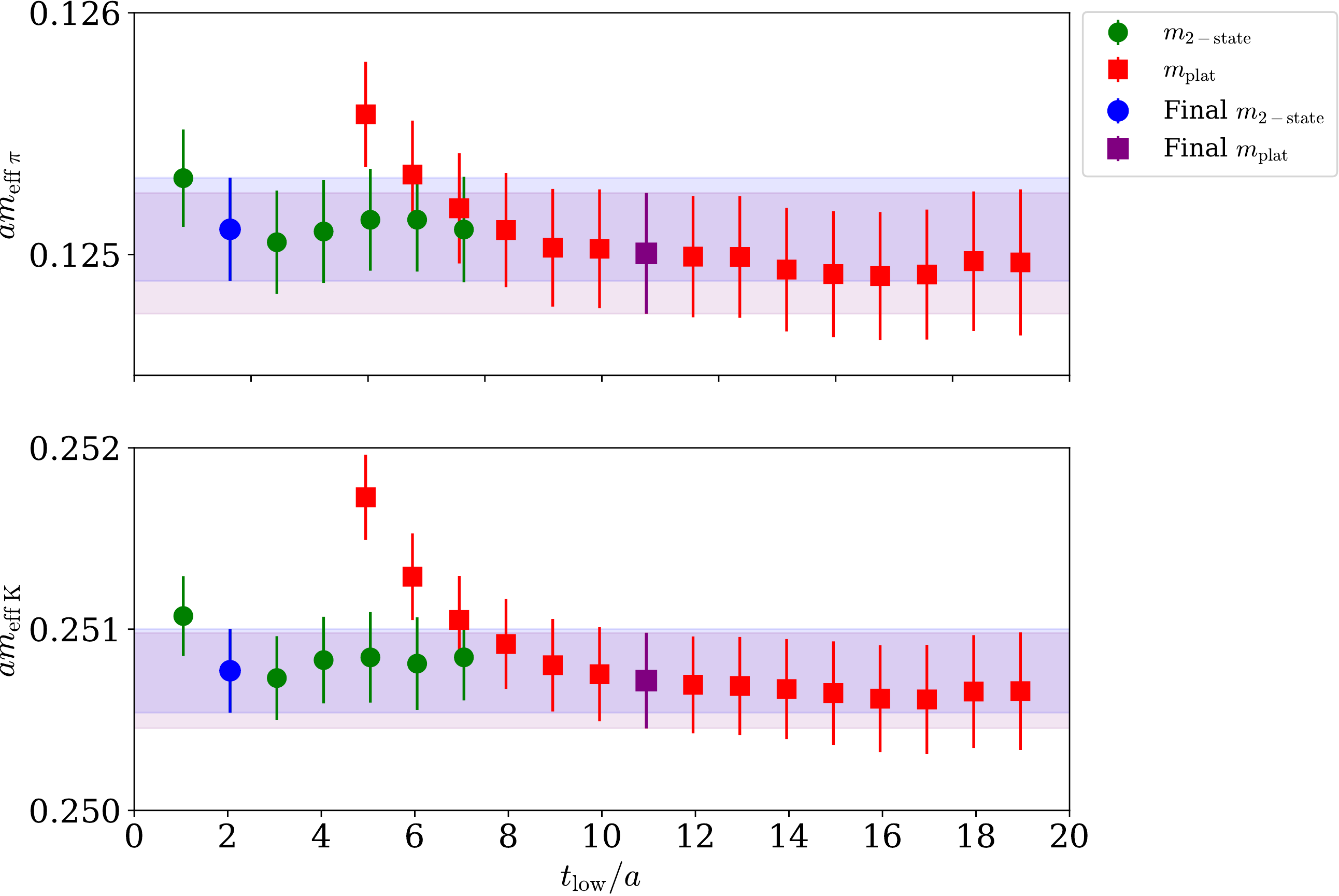}
    \caption{Pion (top) and kaon (bottom) mass in the rest frame as a function of the lowest value of $t/a$, $t_{\rm low}/a$ entering the fit. Results using the plateau method are shown with red squares, and results from the two-state fit with green circles. The selected values extracted using the plateau and two-state fits are shown with the purple square and blue circle, respectively. }
    \label{fig:fit_quality_rest}
\end{figure}

In Fig.~\ref{fig:fit_quality_rest} we show the pion and kaon  mass as a function of the lowest value of $t/a$ entering the fit. We note that for the kaon we  use the so-called Osterwalder Seiler (OS) fermions \cite{OSTERWALDER1978440}
which avoids the mixing effects between strange and charm quarks. The value of the $\mu_s=0.022$ which enters the strange quark propagator is fixed by the physical ratio $m_{D_s}/f_{D_s}=7.9$~\cite{Kostrzewa:2017aom}, where $m_{D_s}$ is the mass of the $D_s$ meson and $f_{D_s}$ it's decay constant. We compare the results extracted from the plateau  and two-state fits of Eq.~(\ref{eq:tsf_twop}). We find that there is a very good agreement between the two methods, when $t_{\rm low}/a \ge 11$ in the plateau fit.

\begin{figure}[h!]
    \centering
    \includegraphics[scale=0.45]{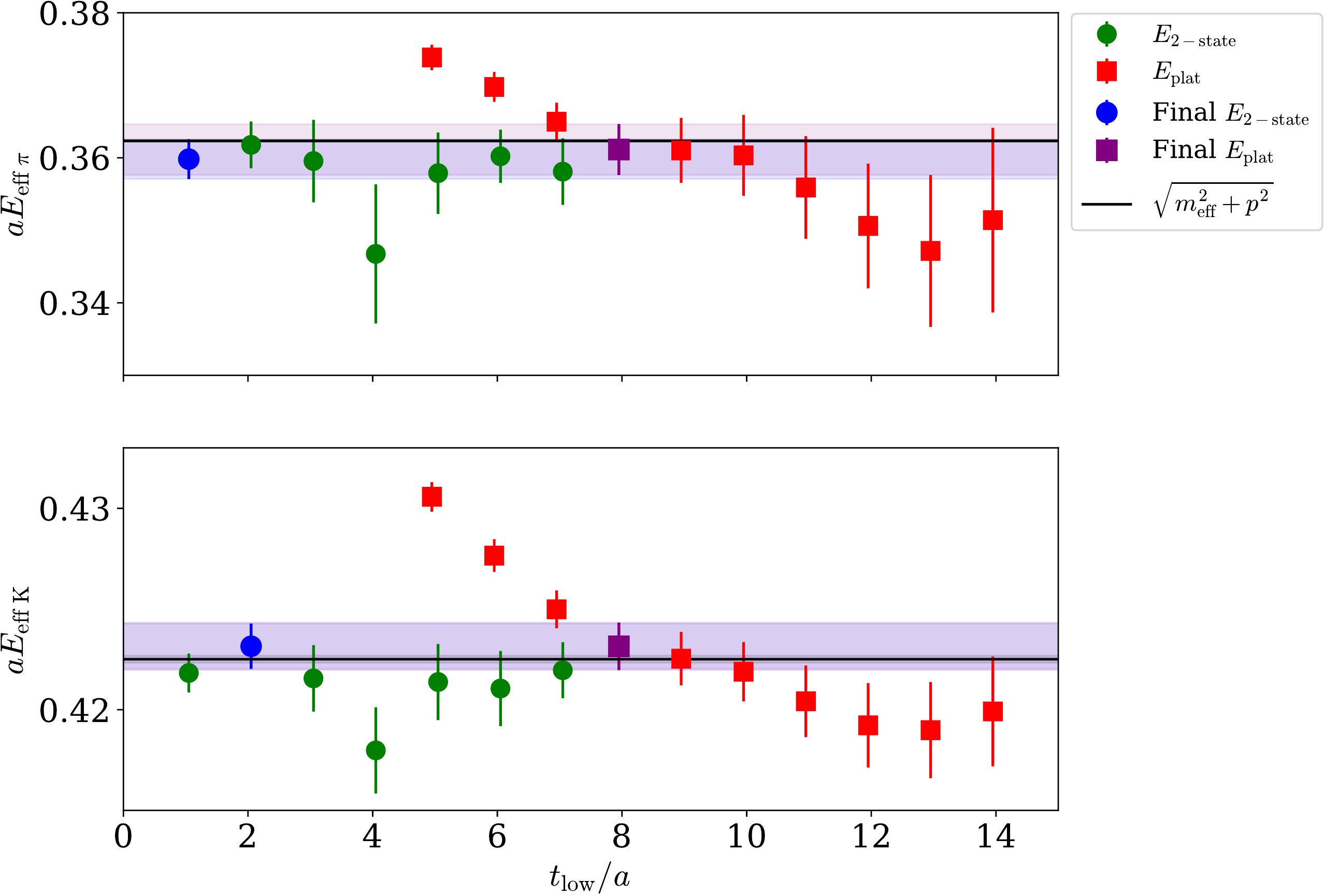}
    \caption{Pion (top) and kaon (bottom) mass in the boosted frame as a function of the the lowest value of $t/a$ entering the fit. The notation is the same as in Fig.~\ref{fig:fit_quality_rest}.}
    \label{fig:fit_quality_boosted}
\end{figure}

We repeat a similar process of extracting the energy and amplitude of the ground state, using the data in the boosted frame. As explained above, we focus on meson momentum boost $\vec{p}_i\,^2=\frac{12\pi^2}{L^2}$. To increase the accuracy of the results and improve the stability of the fit, we perform the various fits on the averaged two-point functions over the eight values of the momentum boost leading to the same $\vec{p}_i\,^2$ ($\vec{p}_i=\frac{2\pi}{L}(\pm1,\pm1,\pm1)$). The results of the fit are shown in Fig.~\ref{fig:fit_quality_boosted}

\medskip
The final values shown with purple and blue in Figs.~\ref{fig:fit_quality_rest} and \ref{fig:fit_quality_boosted} are selected based on the following criterion: We accept a plateau fit with $t_{\rm low}$ when the lowest state mass (energy) in the rest (boosted) frame obtained using the plateau method, $m_{\rm plat}$ ($E_{0,\rm{plat}}$) and the two-state fit $m_{\rm 2st}$ ($E_{0,\rm{2st}}$) satisfy the conditions
\begin{equation}
    \frac{1}{2} \delta m_{\rm plat} \stackrel{>}{=} |m_{\rm plat}-m_{\rm 2st.}|, \quad
    \label{eq:goodFitCondition}
\end{equation}
where $\delta m_{\rm plat}$ is the statistical error on the value extracted from the plateau method.   An additional constraint for the accepted fit, is $\chi_{\rm plat}^2/{\rm d.o.f}<1$. Our final values for the pion mass in the rest frame based on the above criteria are: 
\begin{eqnarray}
{\rm plateau}&:& a m_\pi=0.1250(2) ,\quad t_{\rm low}/a=11 \,,\\
{\rm 2{-}state}&:& a m_\pi=0.1251(2) ,\quad t_{\rm low}/a=2 \,
\end{eqnarray}
while in the boosted frame we obtain
\begin{eqnarray}
{\rm  plateau}&:& a E_{0\pi}=0.361(4) ,\quad t_{\rm low}/a=8 \,,\\
{\rm 2{-}state}&:& a E_{0\pi}=0.360(3) ,\quad t_{\rm low}/a=1\,.
\end{eqnarray}
A similar procedure for the kaon leads to 
\begin{eqnarray}
{\rm plateau}&:& a m_K=0.2507(3),\quad t_{\rm low}/a=11 \,,\\
{\rm 2{-}state}&:& a m_K=0.2508(2),\quad t_{\rm low}/a=2 \,
\end{eqnarray}
in the rest frame, and to
\begin{eqnarray}
{\rm plateau}&:& a E_{0K}=0.423(1) ,\quad t_{\rm low}/a=8 \,,\\
{\rm 2{-}state}&:& a E_{0K}=0.423(1) ,\quad t_{\rm low}/a=2\,
\end{eqnarray}
in the boosted frame.

\begin{figure}[h] 
\hspace*{-1.5cm}
    \includegraphics[scale=0.275]{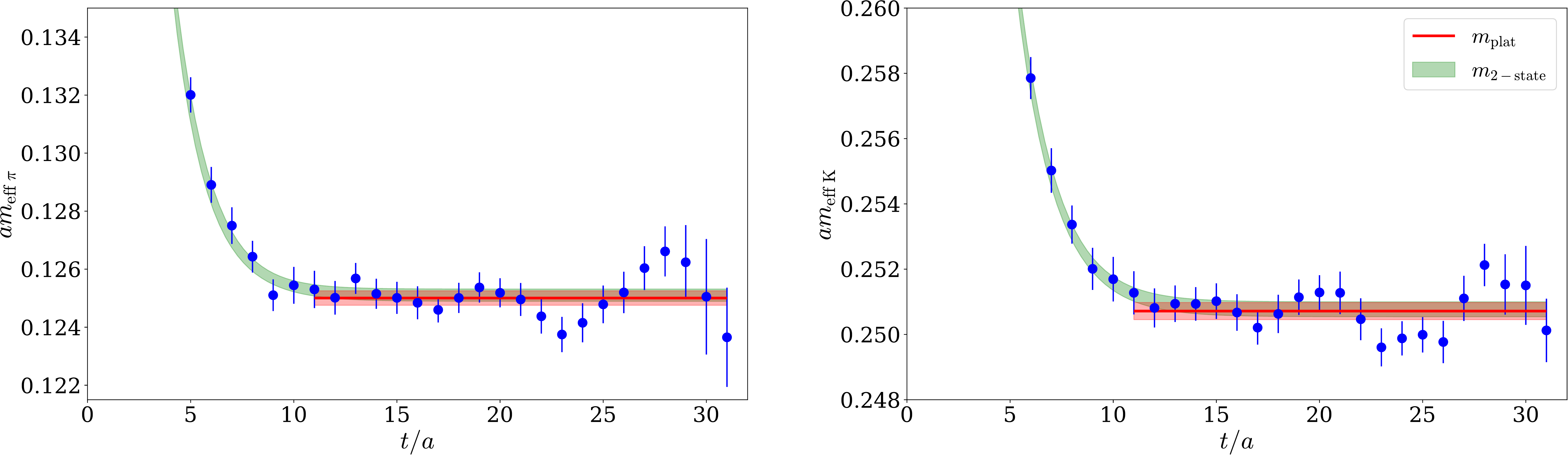}
    \caption{Pion (left) and kaon (right) $\meff$ in the rest frame. The fitted value from the plateau $m_{\rm plat}$  is shown with a red band, and from the two-state fit applied on $\meff$ with a green band.}
    \label{fig:mEff}
\end{figure}

In Fig.~\ref{fig:mEff} we plot the effective mass in the rest frame calculated from Eq.~(\ref{eq:meff}). We also show the plateau fit value and the two-state fit on the correlator as chosen based on Eq.~(\ref{eq:goodFitCondition}). We find full agreement between the two fits, for both the pion  and the kaon.

\subsection{Excited-states contamination in $\avgx$}
\label{sec:excited_states}

To extract the ground state contributions to $\avgx$, one has to ensure suppression of excited states in the three-point functions. This is achieved at sufficiently large insertion ($t/a\gg1$) and sink times ($(t_s-t)/a\gg 1$), where the ground state of the hadron gives the dominant contribution to the three-point correlation functions. It is in this region that we need to extract the matrix elements in order to control excited-states contamination. We employ six values of $t_s$ in the rest frame, $t_s/a \in [12,24]$, which for mesons can be achieved with a reasonable computational cost.  For the pion this is due to the fact that the statistical error for meson matrix elements in the rest frame remains the same with increasing $t_s$. Similarly to the analysis of the two-point functions we use two different analysis methods on the three-point functions, in order to study the convergence to the ground state and the significance of excited-states contributions. We used this study in the rest frame, as a guidance for the $t_s$ values to be employed in the boosted frame. Conclusions from such a study are also useful in a follow-up work for other studies of pion and kaon matrix elements.

\textit{Plateau method:} The first method is based on a constant fit applied to an appropriate ratio of three-point and two-point functions. We choose a convenient ratio so that the denominator contains the ground state obtained from the fit of Eq.~(\ref{eq:tsf_twop}) (instead of the actual two-point functions). This removes the $t_s$ dependence in the ratio, allowing plateau convergence with increasing $t_s$:
\begin{equation}
R_M^{44}(t_s, t)=\frac{C^{44}_M(t_s,t)}{c_0 \,e^{-E_0 t_s}}\,.
    \label{eq:ratio}
\end{equation}
We perform a constant fit as a function of the time $t$ of the operator insertion for each $t_s$ separately and in a region where mild $t$-dependence is observed. {In particular, we use the fit range $t_s/2-2 \le t
\le t_s/2+2$}. One then seeks convergence of the extracted plateau values as $t_s$ increases. In the limit of large time separations the ratio becomes time-independent, that is
\be
R_M^{44}(t_s, t) \xlongrightarrow[\text{$Et\gg1$}]{\text{$\Delta E(t_s-t)\gg1$}}  \Pi^{44}_M\,.
    \label{eq:ratio2}
\ee
Combining Eq.~(\ref{decomp_x_simple}) with Eq.~(\ref{eq:ratio2}), we can obtain $\avgx$ via
\begin{equation}
    \avgx^M = -\frac{4}{3 m_M} {\cal Z}_{\rm vD}\, \Pi^{44}_M\,.
    \label{eq:avgX_plateau}
\end{equation}
In the above expression we include the renormalization function for the one-derivative operator, ${\cal Z}_{\rm vD}$, and all kinematic and normalization factors.

\medskip
\textit{Two-state method:} In the second method of extracting $\avgx$ we take into account the contribution from the first-excited state in the three-point correlation functions. A two-state fit may be performed via
\begin{gather}
    \begin{split}
    C^{44}(t,t_s) = \:
    &A_{00}\left\{ \theta(t_s-t)e^{-E_0t_s} - \theta(t-t_s)e^{-E_0(T-t_s)} \right\} \\
    +& A_{01}\left\{ 
    \theta(t_s-t)e^{-E_0(t_s-t)-E_1t} -
    \theta(t-t_s)e^{-E_0(t-t_s)-E_1(T-t)} \right\} \\
    +& A_{10}\left\{ 
    \theta(t_s-t)e^{-E_1(t_s-t)-E_0t} -
    \theta(t-t_s)e^{-E_1(t-t_s)-E_0(T-t)} \right\} \\
    +& A_{11}\left\{ \theta(t_s-t)e^{-E_1t_s} - \theta(t-t_s)e^{-E_1(T-t_s)} \right\}\,,
    \end{split}
    \label{eq:tsf_threep}
\end{gather}
where $\theta(t)=1$ for $t \ge 0$ and $\theta(t)=0$ for $t<0$. Given the large number of parameters, in Eq.~(\ref{eq:tsf_threep}) we use $m$ for the rest frame and $E_0$ for the boosted frame and $E_1$ for the excited states extracted from the two-state fit  of Eq.~(\ref{eq:tsf_twop}). Therefore, the actual free parameters are the amplitudes $A_{00}$, $A_{01}$, and $A_{11}$ ($A_{01}=A_{10}$ for zero momentum transfer). The desirable matrix element of the ground state is extracted via
\begin{equation}
    \langle M(0)|{\cal O}^{44}|M(0)\rangle = \frac{A_{00}}{c_0}\,,
    \label{eq:twoFitParamRatio}
\end{equation}
where $c_0$ is the coefficient obtained from Eq.~(\ref{eq:tsf_twop}). Eq.~(\ref{eq:twoFitParamRatio}) leads to the following expression for the renormalized $\avgx$
\begin{equation}
    \avgx^M = -\frac{4}{3 m_M} {\cal Z}_{\rm vD}\, \frac{A_{00}}{c_0}\,.
        \label{eq:avgX_2state}
\end{equation}

  \begin{table}[h]
    \begin{tabular}{cccc}
      \hline\hline\\[-2.5ex]
 $t_s/a$ & $\,\,\avgx_{u^+}^{\pi}$ & $\,\,\avgx_u^{k}$ & $\,\,\avgx_s^{k}$ \\[0.5ex]
\hline\\[-2.5ex]
$\,\,\,$12  &$\,\,\,$0.309(3)   &$\,\,\,$0.278(2)   &$\,\,\,$0.339(2)   \\
$\,\,\,$14  &$\,\,\,$0.287(3)   &$\,\,\,$0.264(2)   &$\,\,\,$0.330(2)   \\
$\,\,\,$16  &$\,\,\,$0.275(3)   &$\,\,\,$0.257(2)   &$\,\,\,$0.325(2)   \\
$\,\,\,$18  &$\,\,\,$0.267(3)   &$\,\,\,$0.252(2)   &$\,\,\,$0.322(2)   \\ 
$\,\,\,$20  &$\,\,\,$0.261(4)   &$\,\,\,$0.248(2)   &$\,\,\,$0.319(2)   \\
$\,\,\,$24  &$\,\,\,$0.255(4)   &$\,\,\,$0.244(3)   &$\,\,\,$0.316(2)   \\
2-state (a)     &$\,\,\,$0.261(3)   &$\,\,\,$0.246(2)   &$\,\,\,$0.317(2)   \\ 
2-state (b)     &$\,\,\,$0.262(4)   &$\,\,\,$0.246(2)   &$\,\,\,$0.317(2)   \\ [0.5ex]
      \hline\hline
    \end{tabular}
        \caption{Renormalized data for $\avgx$ for various $t_s$ values and the 2-state fit ((a) $t_s \in [12-24]$, (b) $t_s \in [14-18]$). The numbers shown in the parenthesis are the statistical errors.}
        \label{table:avgx}
  \end{table}
  
In Table~\ref{table:avgx} we collect the values for $\avgx$ for the pion and kaon extracted from different source-sink time separations, and the two-state fit using $t_s/a=12-24$. The results for the two-state fit using $t_s/a=14-18$ is also included. The latter choice is based on investigating the dependence of $\avgx$ on the fit range. We find that the excited-state fit is compatible with the values obtained from $t_s/a \gtrsim 18$ for both the pion and the kaon. As expected in the rest frame, the statistical uncertainties remain constant with increase of the source-sink separation. We find that the plateau values have statistical errors of $2\%$ or less.

\begin{figure}[h!]
    \centering
    \includegraphics[scale=0.48]{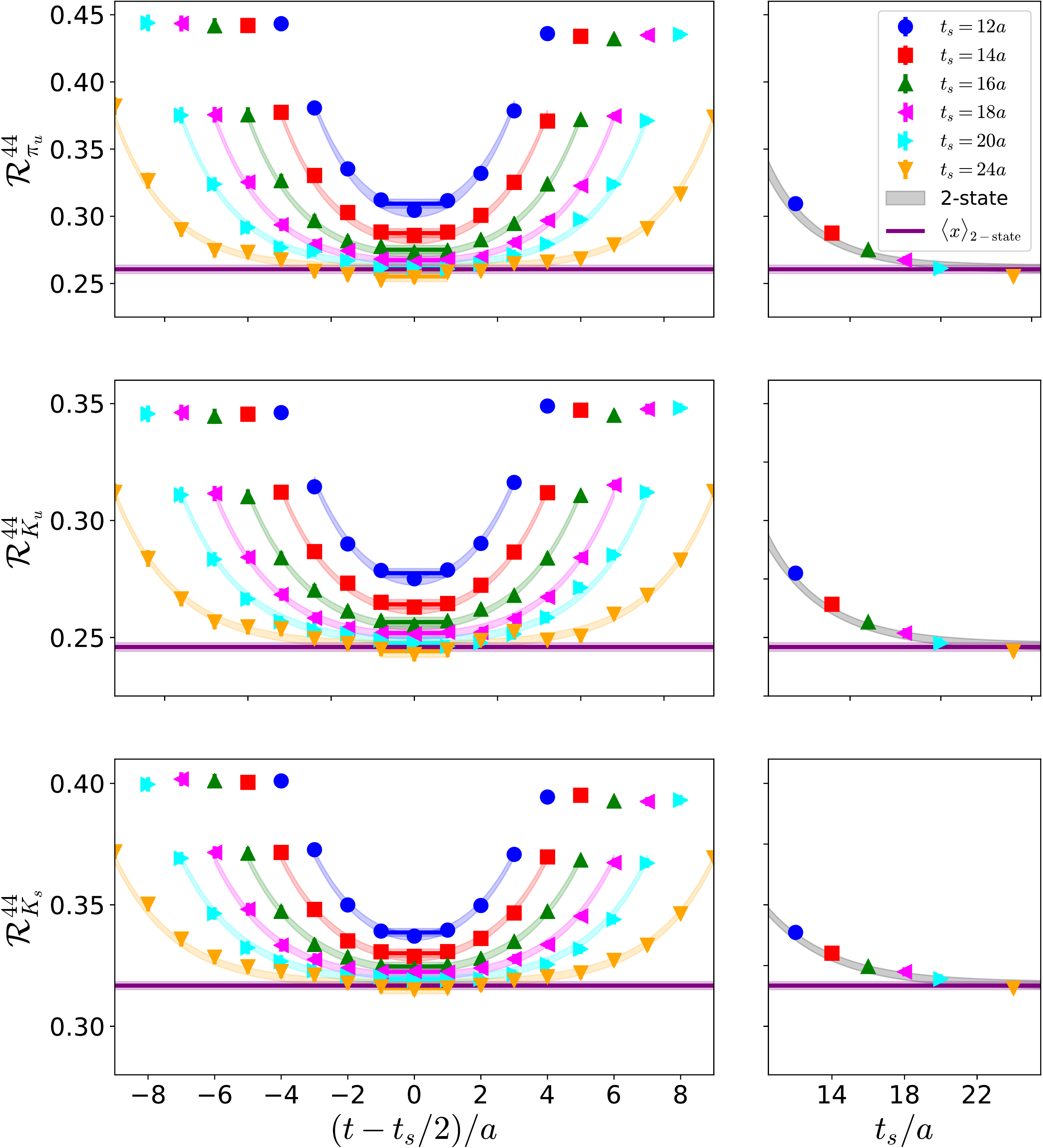}
    \caption{Results for the ratio leading to $\avgx$. We show the up part of the pion, the up and strange parts of the kaon, in the top, center and bottom panels, respectively. In the left column, the points are the ratios in Eq.~(\ref{eq:ratio}) multiplied by the renormalization and all kinematic factors. We plot values for $t_s/a=12$ up to  $t_s/a=24$. The blue, red, green, magenta, cyan and orange constant bands are the results obtained from Eq.~(\ref{eq:avgX_plateau}). The purple band is the two-state fit value obtained from Eq.~(\ref{eq:avgX_2state}). In the right column we plot the plateau values, together with two-state fit results. The gray band is the function obtained from a two-state fit using $t_s \in [12a - 24 a]$ and taking $t=t_s/2$.}
    \label{fig:avgX}
\end{figure}

\newpage
The ratios of three- to two-point functions for  each value of $t_s$ are shown in Fig.~\ref{fig:avgX}, for the up contribution to the pion and the up and strange contribution of the kaon. ${\cal R}$ denotes the ratio $R_M^{44}$ of Eq.~\ref{eq:ratio} multiplied by all kinematic factors and the renormalization function. We observe that the excited-states contamination is similar for both the pion and kaon. We find convergence on $\avgx$ for $t_s \gtrsim 18a$. A comparison of the two-state fit using $t_s \in [12a - 24 a]$ and the plateau values is shown in the right panels of Fig.~\ref{fig:avgX}.
Since the ground-state contribution is established at  $t_s \ge 18a$ and the two-state fits using $t_s \in [12a - 24 a]$ and $t_s \in [14a - 18 a]$ yield the same values, we choose to limit our calculations to $t_s/a=14,\,16,\,18$ for the boosted frame. A comparison for $\avgx$ between the two frames is discussed in Sec.~\ref{sec:x_rest_vs_boosted}.

\subsection{Alternative setup for $\avgx$ in the boosted frame}
\label{sec:x_rest_vs_boosted}

In the above discussion we have used the rest frame for the extraction of $\avgx$, $\vec{p_f}=\vec{0}$, and in the forward limit we also have $\vec{p_i}=\vec{0}$. In this paragraph we explore an alternative setup, a boosted frame with $\vec{p_f}=\vec{p_i}\ne\vec{0}$. Note that employing such a frame is not necessary for $\avgx$, as $A_{20}(0)$ has a nonzero kinematic coefficient in the rest frame. However, the study of $\avgx$ within the boosted frame is interesting because one can understand how the statistical errors increase with $t_s$. Based on the conclusions from Sec.~\ref{sec:excited_states} on the analysis of excited states, we focus on $t_s/a=14,\,16,\,18$, as the computational cost for the same number of configurations is increased by a factor of 8 as compared to the rest frame.
Since this calculation is part of a wider set of operators, the optimal class of momenta is $\vec{p}_i^2=\frac{12\pi^2}{L^2}$. This corresponds to eight combinations for the spatial components, that is, $\vec{p}=\frac{2\pi}{L} (\pm1,\pm1,\pm1)$. In such a frame, the appropriate decomposition is given in Eq.~(\ref{decomp_x_2}), instead of Eq.~(\ref{decomp_x_simple}).

\begin{figure}[h]
    \centering
\hspace*{-0.5cm} 
\includegraphics[scale=0.4]{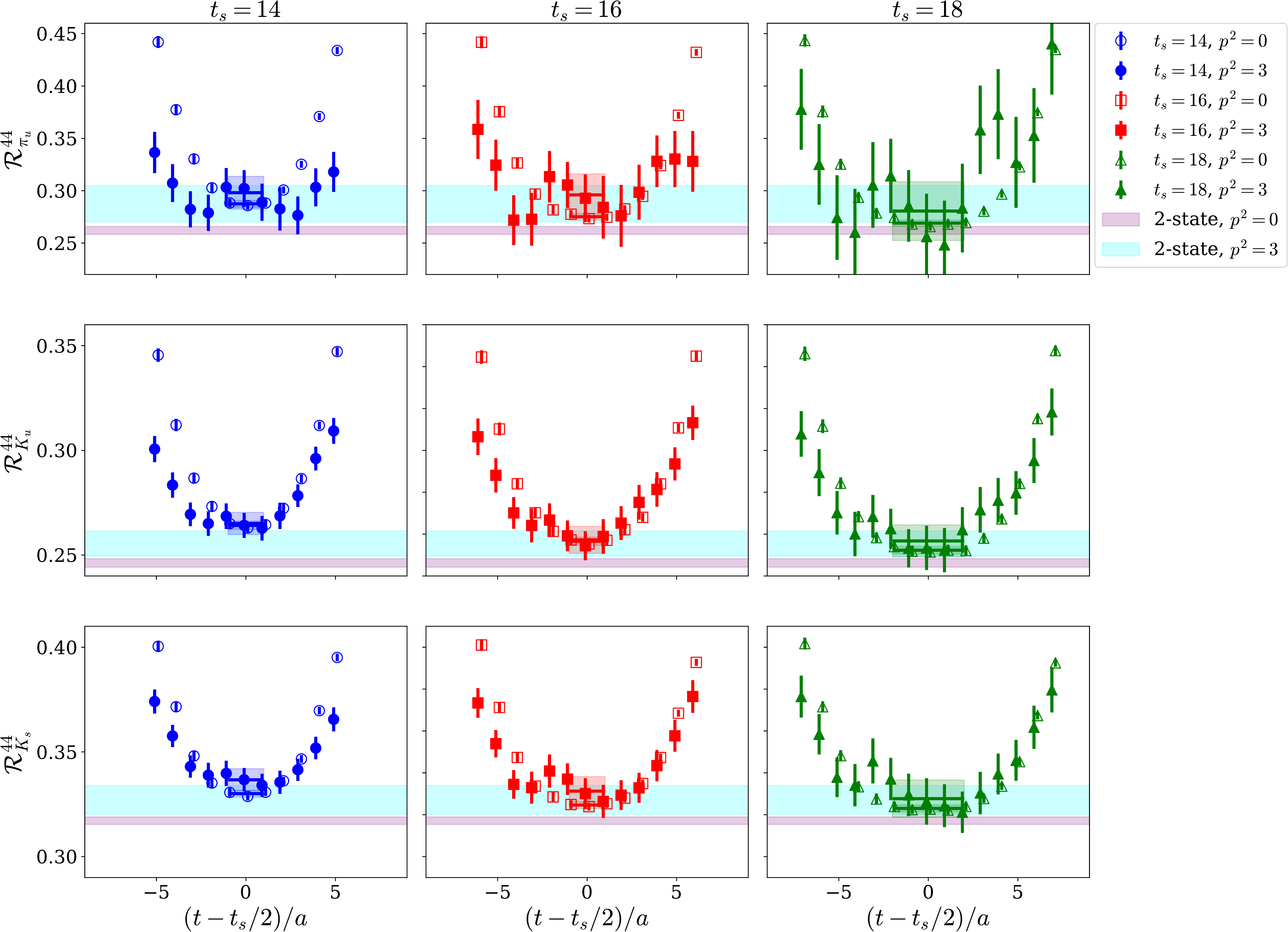}
    \caption{Comparison of $\avgx$ in the boosted (filled symbols) and rest (open symbols) frame. From top to bottom we show  $\avgx$ for the pion and kaon  up and strange contribution. Results at $t_s/a=14,\,16,\,18$ are shown in the left, center and right columns, respectively. For this comparison, we use 16 source positions for each momentum frame so that the statistics are consistent.}
    \label{fig:avgX_psq}
\end{figure}

In Fig.~\ref{fig:avgX_psq} we compare the ratios leading to $\avgx$ for both the pion (top panels) and kaon (center and bottom panels). The left, center and right columns correspond to $t_s=14a,\,16a,\,18a$, respectively. The ratios include all the kinematic factors, and thus can be compared to each other. 
As can be seen, the statistical uncertainties increase with $t_s$ in the boosted frame, which is expected. For both the pion and kaon we find agreement between the plateau values obtained from the two frames within the uncertainties. 

\begin{figure}[h!]
    \centering
    \includegraphics[scale=0.5]{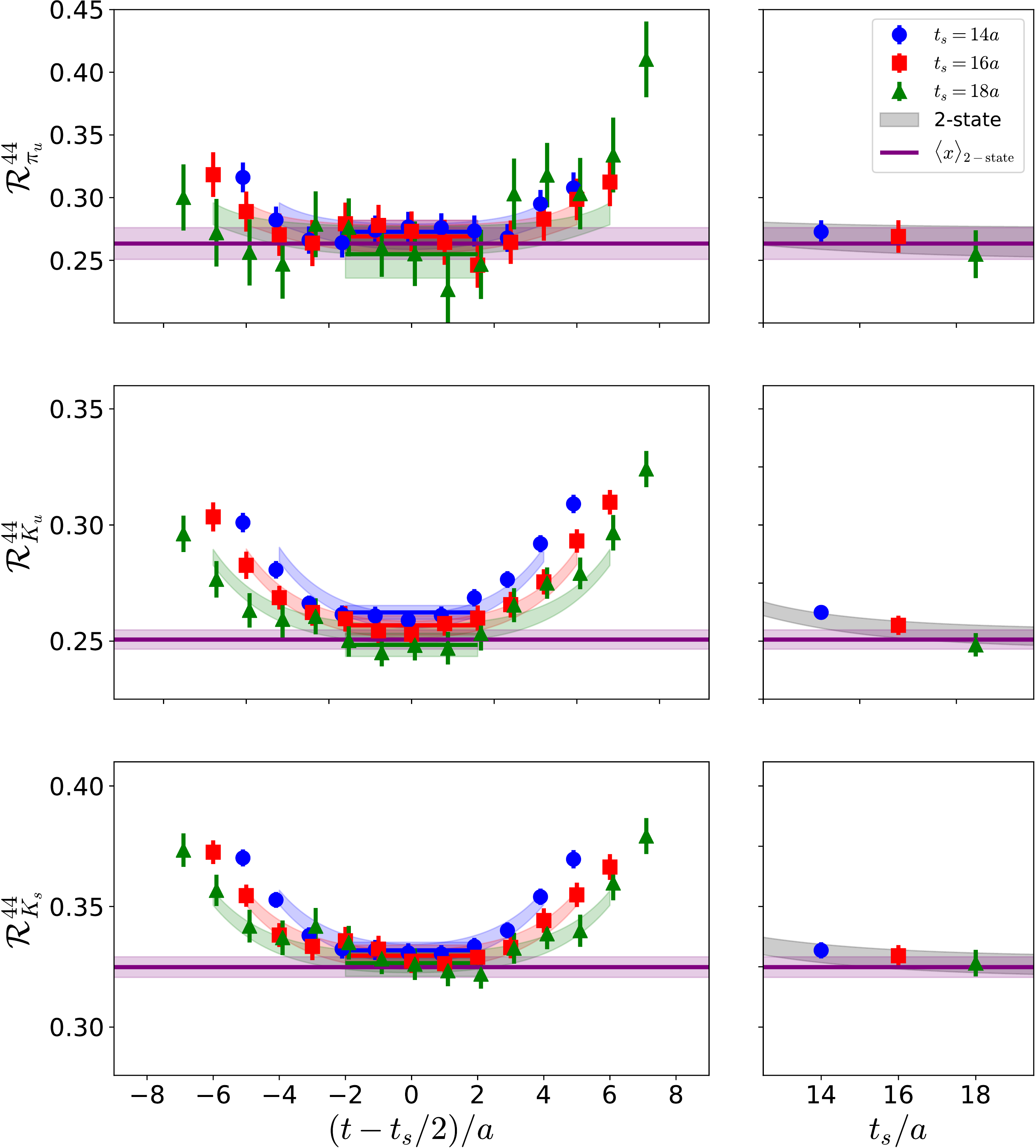}
    \caption{Ratio for $\avgx^\pi_u$ (top), $\avgx^K_u$ (center), and $\avgx^K_s$ (bottom) for $t_s/a=14,\,16,\,18$, shown with blue circles, red squares, and green triangles, respectively. The corresponding plateau values are shown with bands of the same color. The purple band corresponds to the value extracted using the two-state fit.}
    \label{fig:avgX_psq3}
\end{figure}

The ratio for the $\avgx$ for the pion and the kaon is shown in Fig.~\ref{fig:avgX_psq3} for $t_s/a=14,\,16,\,18$, and also compared to the two-state results. We find that all plateau values are compatible with the results of the two-state, indicating that excited-states contamination are within the reported uncertainties, which are larger compared to the ones in the rest frame.
In Table~\ref{table:avgx_boosted} we collect all the results obtained in the boosted frame. We find that the statistical uncertainties in $\avgx_u^{\pi}$ grow from $5\%$ to $10\%$, with the increase of $t_s$ from $14a$ to $18a$. The corresponding increase in $\avgx_u^{k}$ ($\avgx_s^{k}$) is $2\%$ to $3\%$ ($2\%$ to $4\%$). We remind that the error in the rest frame is less or equal to $2\%$, and it is constant for all source-sink separations.

  \begin{table}[h!]
    \begin{tabular}{cccc}
      \hline\hline\\[-2.5ex]
 $t_s/a$ & $\,\,\avgx_u^{\pi}$ & $\,\,\avgx_u^{k}$ & $\,\,\avgx_s^{k}$ \\[0.5ex]
\hline\\[-2.5ex]
$\,\,\,$14  &$\,\,\,$0.273(9)   &$\,\,\,$0.262(3)   &$\,\,\,$0.332(3)   \\
$\,\,\,$16  &$\,\,\,$0.269(13)   &$\,\,\,$0.257(4)   &$\,\,\,$0.330(4)   \\
$\,\,\,$18  &$\,\,\,$0.255(19)   &$\,\,\,$0.248(5)   &$\,\,\,$0.327(6)   \\ 
2-state  &$\,\,\,$0.263(13)   &$\,\,\,$0.251(4)   &$\,\,\,$0.325(4) \\ 
      \hline\hline
    \end{tabular}
        \caption{Renormalized data for $\avgx$ for the three $t_s$ values and the 2-state fit using $t_s \in [14-18]$). The number shown in the parenthesis are the statistical errors.}
        \label{table:avgx_boosted}
  \end{table}

\subsection{Excited-states contamination in $\avgxx$}
\label{sec:excited_states_xx)}

The extraction of  $\avgxx$ is more challenging than $\avgx$ for several reasons. Firstly, $\avgxx$ cannot be extracted in the rest frame due to a vanishing kinematic factor in Eq.~(\ref{decomp_x2_2}). The introduction of momentum in the meson states increases the statistical noise, and in our case, the use of a rather large momentum ($\vec{p}_i^2=\frac{12\pi^2}{L^2}$) worsens the signal even more. Secondly, the presence of two covariant derivatives in the operator contribute to the increase of the gauge noise. Thirdly, having three Dirac indices leads to a more complicated renormalization pattern, and, to completely avoid the mixing with operators of equal or lower dimension, the indices of the operator must be selected different from each other. Here we employ the operator ${\cal O}^{\mu\nu4}$.

\begin{figure}[h!]
    \centering
    \hspace{2cm}
    \includegraphics[scale=0.43]{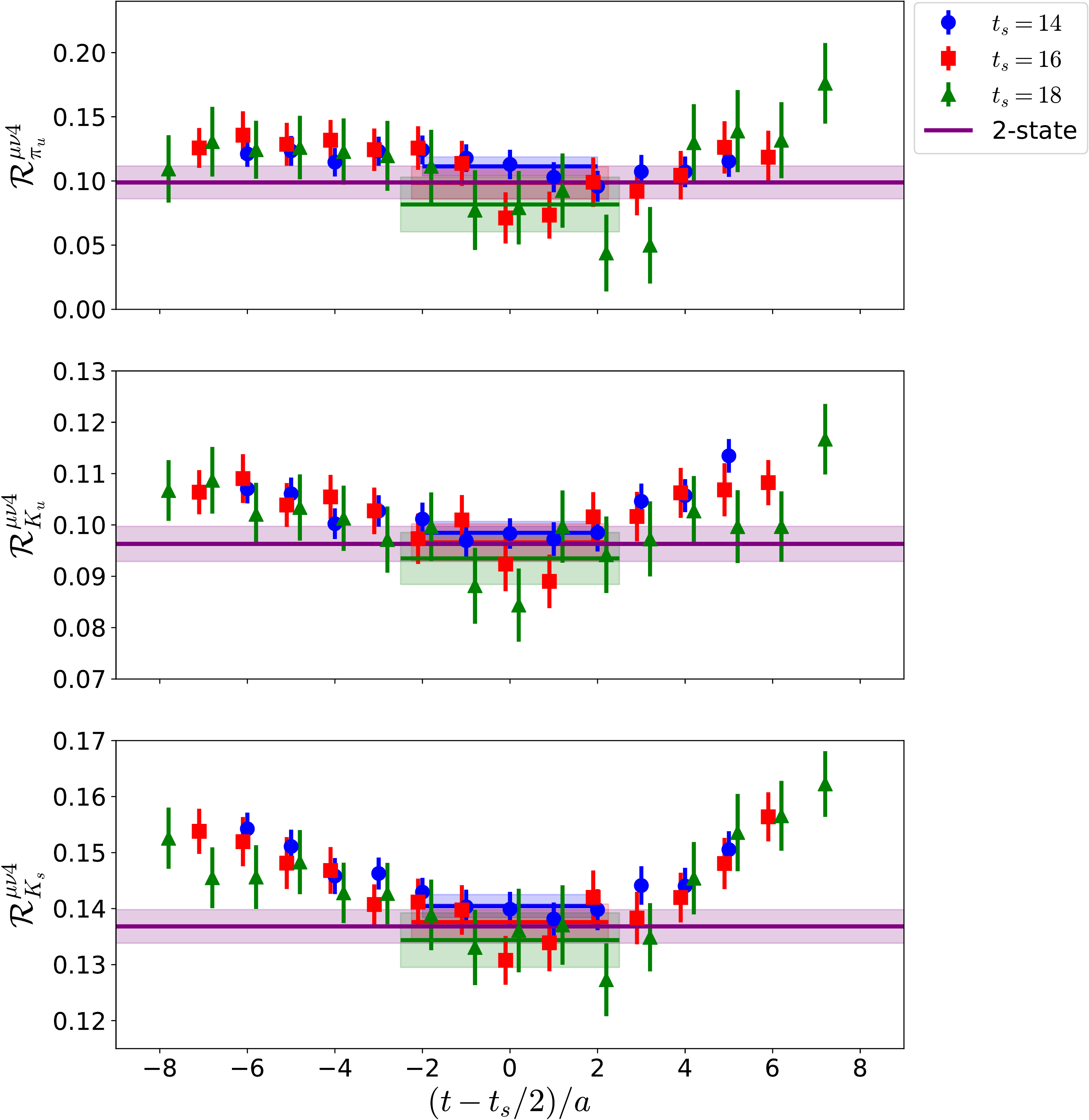}
    \caption{Ratio for $\avgxx^\pi_u$ (top), $\avgxx^K_u$ (center), and $\avgxx^K_s$ (bottom) for $t_s/a=14,\,16,\,18$, shown with blue circles, red squares, and green triangles, respectively. The corresponding plateau values are shown with bands of the same color. The purple band corresponds to the value extracted using the two-state fit.}
    \label{fig:avgX2}
\end{figure}

In Fig.~\ref{fig:avgX2}, we show the ratio leading to $\avgxx$ for the pion  and the kaon. We plot the data for the three values of $t_s$ considered, that is $t_s/a=14,\,16,\,18$, and compare with the two-state results. We find that all plateau values are compatible with the results of the two-state, indicating that excited-states contamination are mild compared to the errors on this quantity. The values obtained from the plateau and two-state fits are given in Table~\ref{table:avgX2}. 

\begin{table}[h!]
    \begin{tabular}{cccc}
      \hline\hline\\[-2.5ex]
 $t_s/a$ & $\,\,\avgxx_u^{\pi}$ & $\,\,\avgxx_u^{K}$ & $\,\,\avgxx_s^{K}$ \\[0.5ex]
\hline\\[-2.5ex]
$\,\,\,$14  &$\,\,\,$0.111(7)   &$\,\,\,$0.098(2)   &$\,\,\,$0.140(2)   \\
$\,\,\,$16  &$\,\,\,$0.099(13)   &$\,\,\,$0.097(4)   &$\,\,\,$0.138(3)   \\
$\,\,\,$18  &$\,\,\,$0.082(21)   &$\,\,\,$0.093(5)   &$\,\,\,$0.134(5)   \\
2-state     &$\,\,\,$0.099(13)   &$\,\,\,$0.096(3)   &$\,\,\,$0.137(3)    \\ [0.5ex]
      \hline\hline
    \end{tabular}
        \caption{Renormalized data for $\avgxx$ at each $t_s$ values and the two-state fit ($t_s \in [14-18]$). The number shown in the parenthesis is statistical error.}
        \label{table:avgX2}
  \end{table}

\section{Final results and comparison with other studies}
\label{sec5}

In this section we discuss our final values for the quantities studied in this work. For $\avgx$ we give the results in the rest frame and using the two-state fits. This is because the statistical uncertainties are the same for all values of $t_s$. For $\avgxx$ we use the results extracted from $t_s/a=18$,  as the two-state fit may be driven by the most accurate data.
\bea
\label{eq:x_pi_fin}
\avgx^\pi_{u^+} &=& 0.261(3)(6)\,,\\[0.5ex]
\label{eq:x_k_u_fin}
\avgx^K_{u^+} &=& 0.246(2)(2)\,,\\[0.5ex]
\label{eq:x_k_s_fin}
\avgx^K_{s^+} &=& 0.317(2)(1)\,,
\eea
and
\bea
\label{eq:x2_pi_fin}
\avgxx^\pi_{u^+} &=& 0.082(21)(17)\,,\\[0.5ex]
\label{eq:x2_k_u_fin}
\avgxx^K_{u^+} &=& 0.093(5)(3)\,,\\[0.5ex]
\label{eq:x2_k_s_fin}
\avgxx^K_{s^+} &=& 0.134(5)(2)\,.
\eea
These results are in the $\overline{\rm MS}$ scheme at a scale of 2 GeV.
We use the notation $q^+ \equiv q+\bar{q}$ for the sum from quark and antiquark contributions. As already mentioned $\avgx^\pi_{u^+ + {d}^+}=2 \avgx^\pi_{u^+}$ and $\avgxx^\pi_{u^+ + {d}^+}=2\avgxx^\pi_{u^+}$.
The number given in the first parenthesis is the statistical error obtained from a jackknife analysis. We also report a systematic error, given in second parenthesis, which is due to excited states contamination. {This systematic error is the difference between the value extracted using the two-state fit and the value of the plateau method at $t_s=24$ for $\avgx$, and $t_s=18$ for $\avgxx$}. We also extract the ratio $\avgxx/\avgx$, for which we find
\bea
\label{eq:x_over_x2_pi_fin}
\frac{\avgxx^\pi_{u^+}}{\avgx^\pi_{u^+}} &=& 0.30(8)(7)\,,\\[1ex]
\label{eq:x_over_x2_k_u_fin}
\frac{\avgxx^K_{u^+}}{\avgx^K_{u^+}} &=& 0.37(2)(2)\,,\\[1ex]
\label{eq:x_over_x2_k_s_fin}
\frac{\avgxx^K_{s^+}}{\avgx^K_{s^+}} &=& 0.42(2)(2)\,,
\eea
using the results at $t_s/a=18$ for $\avgx$ and $\avgxx$. The number in the first parenthesis is statistical, while in the second parenthesis is systematic due to excited states. 

There is very limited experimental data on the kaon PDF, so it is interesting to contrast these moment results with expectations from model calculations. For our lattice results we find $\avgx^K_{u^+} < \avgx^\pi_{u^+} < \avgx^K_{s^+}$ which is consistent with many phenomenological calculations, including the DSE results of Ref.~\cite{Bednar:2018mtf}. This ordering in the momentum fractions is understood because the heavier $s$ quark skews $s_K(x)$ to larger $x$ which is compensated by a shift in $u_K(x)$ to smaller $x$. In the limit of equal quark masses these moments would be degenerate, therefore, we find flavor breaking effects of up to 20\% in these moments. For the third Mellin moment we find $\avgxx^K_{u^+}, \avgxx^\pi_{u^+} < \avgx^K_{s^+}$, which is again consistent with expectations. Uncertainties on the third Mellin for the $u$ quark in the pion and kaon do not allow an ordering of these moments, however, any deviation from the order found for the momentum fractions would be very interesting.

There are a number of calculations on the pion $\avgx$~\cite{Brommel:2006zz,Brommel:2007zz,Baron:2007ti,Bali:2013gya,Oehm:2018jvm}, including results obtained directly at the physical point~\cite{Abdel-Rehim:2015owa}. The pion third Mellin moment $\avgxx$, on the other hand, is lesser known, and has been studied in Refs.~\cite{Brommel:2006zz,Brommel:2007zz,Oehm:2018jvm} using different lattice formulations.  It is worth mentioning that moments of PDFs for the pion and kaon have been extracted using non-local operators~\cite{Karpie:2018zaz,Sufian:2019bol,Izubuchi:2019lyk,Joo:2019bzr,Bali:2019dqc,Lin:2020ssv,Gao:2020ito}. However, we do not attempt comparison with such studies, as they suffer from very different systematic uncertainties. Here we compare with lattice results on $\avgx^\pi_{u^+}$ extracted at the same or similar value of the pion mass, that is 240 - 270 MeV.

In Ref.~\cite{Oehm:2018jvm}, several $N_f=2+1+1$ ensembles of twisted mass fermions with no clover improvement were used for the calculation of the pion moments. They find $\avgx^\pi_{u^+}= 0.2586(41)(28)$ on an ensemble (A30.32) with the similar lattice spacing  ($a=0.09$~fm) and lattice size  to the one of this work. For another ensemble (B25.32) with a smaller lattice spacing $a=0.082$ fm, $m_\pi=260$ MeV and $m_\pi L=3.5$, they found $\avgx^\pi_{u^+}= 0.2523(51)(71)$. Both values are in agreement with $\avgx^\pi_{u^+}=0.261(3)(6)$ obtained using our $N_f=2+1+1$ clover-improved ensemble.

It is interesting to compare with phenomenological estimates, which can be found in Refs.~\cite{Barry:2018ort} and~\cite{Wijesooriya:2005ir}. The older analysis of Ref.~\cite{Wijesooriya:2005ir} gives a value of $\avgx^\pi_{u}=0.217(10)$, in the $\overline{\rm MS}$ at a scale at $(5.2\, {\rm GeV})^2$, while ours is $(2\, {\rm GeV})^2$. Converting to 2 GeV, their value becomes $\avgx^\pi_{u}=0.361(17)$.
A more recent analysis is presented by the JAM Collaboration~\cite{Barry:2018ort}, on a large set of experimental data including Drell-Yan data and leading neutron electroproduction from HERA.  They find $\avgx^\pi_{valence}=0.480(10)$, which is reasonably close to  our value of $2\avgx^\pi_{u^+}=0.522(13)$. The error in the parenthesis is the combined statistical and systematic uncertainties added in quadrature. The fact that our value is  higher, by $\sim 4\%$, maybe attributed to the fact that our calculation is not at the physical point and the continuum limit is not taken. Both the chiral extrapolation and taking $a\rightarrow 0$  will decrease this value as demonstrated in Ref.~\cite{Oehm:2018jvm}. We also note that all lattice calculations to date consider only the connected contributions as done in this work. The disconnected contributions should be included for a final comparison with phenomenology. We summarize the results for $\avgx^\pi_{u}$ in Table~\ref{tab:x_pi_compare}.

There are very limited calculations for $\avgxx$ within lattice QCD, and the one which we can directly compare with our results is Ref.~\cite{Oehm:2018jvm}. They find $\avgxx^\pi_{u^+}=0.131(18)(24)$ and $\avgxx^\pi_{u^+}=0.132(40)(53)$ for ensembles A30.32 and B25.32, respectively. These estimates are compatible with our final value, within the large uncertainties of the aforementioned values. $\avgxx^\pi$ was also calculated in Refs.~\cite{Brommel:2006zz,Brommel:2007zz} using a different operator, which has two Dirac indices the same. Such a choice is expected to lead to more complicated renormalization pattern due to mixing, which is not addressed in Refs.~\cite{Brommel:2006zz,Brommel:2007zz}. They obtain $\avgxx^\pi_{u^+}=0.128(9)(4)$ which is, however consistent with the value obtained in this work.

Phenomenological estimates for $\avgxx^\pi$ can be found in Ref.~\cite{Barry:2018ort} where  a value of $\avgxx^\pi_{valence}=0.210(5)$ is reported, which is compatible with our value of $2\avgxx^\pi_{u^+}=0.164(54)$ within uncertainties. However, one needs to bear in mind that the phenomenological value does not include sea quark contributions unlike the lattice QCD calculation, where such sea quark effects are automatically included.
For completeness, we also provide the results from Ref.~\cite{Wijesooriya:2005ir}, which correspond to a scale of $(5.2 \,{\rm GeV})^2$. Their finding is $\avgxx^\pi_u=0.087(5)$. We convert this result to 2 GeV resulting to $\avgxx^\pi_u=0.169(10)$. This
is compatible with our value.

  \begin{table}[h]
    \begin{tabular}{|l|l|l|}
\hline\\[-2.5ex]
$\qquad$ Reference & $\qquad$$\avgx^{\pi}$ & $\qquad$$\avgxx^{\pi}$ \\[0.5ex]
\hline\\[-2.5ex]
\,\,This work    (lattice)                       &\,\,\,0.522(13)\,\,\,    &\,\,\,0.164(54)\,\,\,   \\
\,\,Ref.~\cite{Oehm:2018jvm} (lattice)           &\,\,\,0.517(99)\,\,\,    &\,\,\,0.262(60)\,\,\,   \\
\,\,Ref.~\cite{Oehm:2018jvm} (lattice)           &\,\,\,0.505 (174) \,\,\, &\,\,\,0.264(133)\,\,\, \\
\,\,Ref.~\cite{Wijesooriya:2005ir} (global fits) &\,\,\,0.361(17) \,\,\,   &\,\,\,0.169(10)\,\,\,   \\
\,\,Ref.~\cite{Barry:2018ort} (global fits)\,\,  &\,\,\,0.480(10) \,\,\,   &\,\,\,0.210(5) \,\,\,   \\[0.5ex]
\hline
    \end{tabular}
        \caption{Comparison of lattice results and phenomenological data for $\avgx^\pi$ and $\avgxx^\pi$.}
        \label{tab:x_pi_compare}
  \end{table}


\section{Summary}
\label{sec6}

We present a calculation of the second and third Mellin moments, $\avgx$ and $\avgxx$ for the pion and kaon. We use one $N_f=2+1+1$ ensemble reproducing a pion mass of 260 MeV and a kaon mass of 530 MeV. For $\avgx$ we employ both the rest and boosted frames, and we find full agreement between the two. However, the statistical uncertainties for the boosted frame are larger, as can be seen in Fig.~\ref{fig:avgX_psq}. To extract $\avgxx$ one requires a boosted frame due to kinematical factors. The selected meson momentum boost has all spatial components nonzero, and gives $|\vec{p}_i|=\frac{\sqrt{12}\pi}{L}$ (0.72 GeV). We renormalize all matrix elements with non-perturbative renormalization with cut-off subtraction that utilizes lattice QCD perturbation theory.

We perform a thorough investigation of excited states using the three-point function that determines $\avgx$. For this investigation we use the rest frame and calculated the matrix elements for six values of the source-sink time separation ranging from 1.12~fm to 2.23~fm. The computational cost for this study is within reach, as the statistical error does not increase with $t_s$ in the rest frame for the pion and only increases mildly for the kaon. We analyze the data using one-state and two-state fits. We find that excited states are suppressed for $t_s>1.6$ fm, that is, $t_s=18 a$ or higher. Another important conclusion from the analysis is that the two-state fits using only $t_s/a=14,\,16,\,18$ are compatible with the two-state fits obtained including the larger $t_s$ values. This is crucial, as in the case of the boosted frame, the statistical errors  increase with $t_s$, as illustrated in e.g, Fig.~\ref{fig:avgX2}, limiting how large $t_s$ can be. Thus, for the boosted frame we perform the computation for $t_s/a=14,\,16,\,18$, where the consistency of the results extracted between one- and two-state fits demonstrates that excited states are correctly accounted for. 

The results for the pion are given in Eq.~(\ref{eq:x_pi_fin}) and Eq.~(\ref{eq:x2_pi_fin}), in the $\overline{\rm MS}$ scheme at a renormalization scale of 2 GeV. Our results agree very well with the lattice QCD analysis of Ref.~\cite{Oehm:2018jvm}. 
It is important to emphasize that the in-depth study and elimination of excited-states in our analysis has reduced the extracted values bringing them closer to those determined from phenomenology. For example, our lattice data for source-sink time separations below 1.6 fm give a value that is $10\% - 20\%$ higher than the final value extracted when the larger  separations are used (see Table~\ref{table:avgx}).

Our final results for $\avgx^K_{u,s}$ and $\avgxx^K_{u,s}$ are given in Eqs.~(\ref{eq:x_k_u_fin}) - (\ref{eq:x_k_s_fin}) and Eqs.~(\ref{eq:x2_k_u_fin}) - (\ref{eq:x2_k_s_fin}), respectively in the $\overline{\rm MS}$ scheme at a scale of 2~GeV. Currently, there are no other lattice data for these quantities, nor global fits on experimental data. Therefore, the results on the kaon presented in this work provide a first prediction. Taking in to account that for the ensemble employed in this work the kaon mass is about 530 MeV, i.e. only $\sim7\%$ heavier than its physical value,  means that the values for $\avgx^K_{u,s}$ and $\avgxx^K_{u,s}$ can be considered as a good approximation of their physical counterparts.

In the near future, we will consider calculation of $\langle x^3 \rangle$ for both the pion and kaon. Another direction is the form factors and generalized form factors, which require off-forward matrix elements. We intend to test the momentum smearing method~\cite{Bali:2016lva} for the boosted frame, which has been proven to increase the overlap with the ground state, decreasing significantly the statistical noise. This method is successful in the reduction of statistical noise in hadron matrix elements of non-local operators (see, e.g., Refs.~\cite{Alexandrou:2016jqi,Alexandrou:2018pbm,Alexandrou:2018eet,Karpie:2018zaz,Alexandrou:2019lfo,Joo:2019jct,Joo:2020spy,Bhattacharya:2020cen,Bhat:2020ktg,Gao:2020ito}).

\begin{acknowledgements}

We would like to thank all members of ETMC for a very constructive and enjoyable collaboration. We would also like to thank the JAM Collaboration for providing their updated results for the pion moments.
M.C. and C.L. acknowledges financial support by the U.S. National Science Foundation under Grant No.\ PHY-1714407. 
K.H. is supported by the Cyprus Research Promotion foundation under contract number POST-DOC/0718/0100.   This project has received funding from the Horizon 2020 research and innovation program
of the European Commission (EC) under the Marie Sk\l{}odowska-Curie grant agreement No 642069. S.B. is partially supported by this program as well as from the project  COMPLEMENTARY/0916/0015 funded by the Cyprus Research Promotion Foundation and from the EC H2020 infrastructure project PRACE-6IP grant agreement No 823767.
C.L. acknowledges support from the U.S. Department of Energy, Office of Science, Office of Nuclear Physics, contract no. DE-AC02-06CH11357. C.A. acknowledges support by the internal program of  the University of Cyprus {\it Nucleon parton distribution functions using Lattice Quantum Chromodynamics}.
This project was funded in part by the DFG as a project in the Sino-German CRC110.
This work used computational resources from Extreme Science and Engineering Discovery Environment (XSEDE), which is supported by National Science Foundation grant number TG-PHY170022. 
This research includes calculations carried out on the HPC resources of Temple University, supported in part by the National Science Foundation through major research instrumentation grant number 1625061 and by the US Army Research Laboratory under contract number W911NF-16-2-0189. 
This work was in part supported by the U.S. Department of Energy, Office of Science, Office of Nuclear Physics, contract no.~DE-AC02-06CH11357.

\end{acknowledgements}

\bibliography{avgX_mesons.bib}

\end{document}